\documentclass[manuscript]{acmart}

\usepackage{algorithmic}
\usepackage{algorithm}
\usepackage{subcaption}
\usepackage{multirow}

\usepackage{pifont} 

\usepackage{tcolorbox} 

\AtBeginDocument{%
  \providecommand\BibTeX{{%
    \normalfont B\kern-0.5em{\scshape i\kern-0.25em b}\kern-0.8em\TeX}}}

\setcopyright{acmcopyright}
\copyrightyear{2023}
\acmYear{2023}

\setcopyright{rightsretained}

\begin{document}

\title{Stopping Methods for Technology Assisted Reviews based on Point Processes}

\author[1]{Mark Stevenson}
\email{mark.stevenson@sheffield.ac.uk}
\orcid{0000-0002-9483-6006}
\affiliation{%
  \institution{University of Sheffield}
\streetaddress{Regent Court, 211 Portobello}
\city{Sheffield}
\state{S Yorks}
\country{United Kingdom}
}

\author[1]{Reem Bin-Hezam}
\email{rybinhezam@pnu.edu.sa}
\orcid{0000-0002-9156-6186}
\affiliation{%
  \institution{University of Sheffield}
\streetaddress{Regent Court, 211 Portobello}
\city{Sheffield}
\country{United Kingdom}
}
\affiliation{%
  \institution{Princess Nourah Bint Abdulrahman University}
\streetaddress{Department of Information Systems, College of Computer and Information Sciences}
\country{Saudi Arabia}
}

\renewcommand{\shortauthors}{Stevenson and Bin-Hezam}

\begin{abstract}
Technology Assisted Review (TAR), which aims to reduce the effort required to screen collections of documents for relevance, is used to develop systematic reviews of medical evidence and identify documents that must be disclosed in response to legal proceedings.
Stopping methods are algorithms which determine when to stop screening documents during the TAR process, helping to ensure that workload is minimised while still achieving a high level of recall. This paper proposes a novel stopping method based on point processes, which are statistical models that can be used to represent the occurrence of random events. The approach uses rate functions to model the occurrence of relevant documents in the ranking and compares four candidates, including one that has not previously been used for this purpose (hyperbolic). Evaluation is carried out using standard datasets (CLEF e-Health, TREC Total Recall, TREC Legal), and this work is the first to explore stopping method robustness by reporting performance on a range of rankings of varying effectiveness. Results show that the proposed method achieves the desired level of recall without requiring an excessive number of documents to be examined in the majority of cases and also compares well against multiple alternative approaches. 
\end{abstract}


\acmJournal{TOIS}
\acmYear{2023} \acmVolume{1} \acmNumber{1} \acmArticle{1} \acmMonth{1} \acmPrice{}\acmDOI{10.1145/3631990}


\maketitle

\section{Introduction} 

Technology Assisted Review (TAR) aims to minimise the manual effort required to screen a collection of documents for relevance. Applications of TAR include scenarios in which the aim is to retrieve as many documents that meet an information need as possible, and preferably all documents. For example, in systematic reviewing, a key foundation of evidence-based medicine that is also common in other fields, research questions are answered based on information from the scientific literature \cite{higgins2019cochrane}. The standard approach to identifying relevant literature is to construct a Boolean query designed to optimise recall over precision and manually screen the results, an expensive and time-consuming process that can involve manual assessment of tens of thousands of documents \cite{shemilt2016use,michelson2019significant}.  In the legal domain, electronic discovery (eDiscovery) is the identification of documents for legal purposes, such as disclosure in response to litigation \cite{roegiest2015trec,grossman2016trec,oard2018jointly} or to meet the requirements of freedom of information (FoI) legislation \cite{mcdonald2020accuracy,baron2020providing}. In eDiscovery, it is important to identify as many relevant documents as possible given the resources available to ensure compliance with legal obligations and avoid potential penalties. Identifying relevant information in response to FoI requests ensures that sensitive information is not released inadvertently. In Information Retrieval (IR), test collections are a key component of the standard evaluation methodology. Maximising the number of relevant documents identified reduces the potential for bias when evaluating retrieval models \cite{cormack1998efficient} but becomes more difficult to achieve as increasing volumes of information become available in electronic format and the size of these collections increases. 

Approaches to the TAR problem generally focus on the development of efficient ranking approaches that aim to rank relevant documents highly, thereby ensuring that they are discovered as early as possible. Continuous Active Learning (CAL) has proved to be a successful version of this approach \cite{cormack2014evaluation,cormack2015autonomy,cormack2016scalability,li2020stop}. CAL relies on a classifier to rank documents in the collection. Initial training of the classifier can be achieved in various ways such as using a small number of relevant documents (often referred to as ``seeds") or using the query as a pseudo-document. The classifier is then applied to the document collection and some portion of the documents examined. The relevance judgements produced by this process are used to re-train the classifier which is then used to re-rank the remaining documents. The classifier's accuracy improves as the process is repeated which leads to the relevant documents being identified early in the ranking. 

However, even the most effective document ranking does not reduce reviewer workload if they are still required to screen all documents in the collection. A key problem within TAR is therefore deciding when a reviewer can stop examining documents \cite{cormack2016engineering}. This leads to the need for effective stopping methods which reviewers can combine with ranking approaches, such as CAL, to inform their decision about whether to stop examining documents. 

A reviewer's {\it target recall} for a TAR problem is the minimum percentage of relevant documents in a collection that they aim to identify before they cease examining documents. Following \citet{cormack2016engineering}, a TAR stopping method is a mechanism to predict when a reviewer has examined a sufficient number of documents to achieve the target recall while also minimising the total number of documents examined. 

A range of stopping methods has been proposed in the literature (see Section \ref{sec:background} for a more complete review). The simplest of these are based on ad-hoc methods to identify the point in the ranking where the target recall has been reached, e.g. \cite{ros2017machine,Losada2019}. These approaches do not provide the reviewer with any indication of confidence in their decision and often rely heavily on parameters being set to appropriate values. Another, more common, approach is to attempt to estimate the total number of relevant documents in the collection and inform the user when they have reached the target recall. This may be achieved by training a classifier (such as one developed for a CAL-type approach) and using it to estimate the number of relevant documents in the unexamined portion of the ranking, e.g. \cite{Howard2020,callaghan2020statistical,Yang2021heuristic}. However, these approaches generally assume that the rate at which documents occur in the unexamined portion is the same as the portion that has been observed, which is unlikely to be the case for any reasonable document ranking. 

This paper builds on previous work on TAR stopping methods, particularly \citet{cormack2016engineering} and \citet{li2020stop}, to develop a novel approach based on point processes \cite{Sneyd19,Sneyd21}. Point processes are well understood statistical models which use information about the rate at which relevant documents are observed in a ranking to make inferences about the total number in (part of) a collection. They have the advantage of being able to model the fact that, in any reasonable ranking, relevant documents are more likely to appear early in the ranking. This paper develops a stopping method based on two types of point processes: Poisson Process and Cox Process. It also compares four approaches to modelling the rate at which relevant documents occur, including three that have been used in previous work on stopping methods and one which has not. 
These methods are evaluated and compared against alternative approaches on a range of datasets used to evaluate TAR approaches: the CLEF Technology-Assisted Review in Empirical Medicine \cite{kanoulas2017clef,kanoulas2018clef,kanoulas2019clef}, the TREC Total Recall tasks \cite{grossman2016trec}  and the TREC Legal Tasks \cite{cormack2010overview}. These experiments include evaluation using the complete set of runs submitted for the CLEF Technology-Assisted Review in Empirical Medicine dataset, allowing the assessment of their effectiveness over rankings of varying effectiveness.\footnote{Code to reproduce the experiments reported in this paper is available from \url{https://github.com/ReemBinHezam/TAR_Stopping_Point_Processes}}

The contributions of this work can be summarised as follows: 
\vspace{-1eX}
\begin{itemize}
    \item Proposes a novel stopping method for TAR based on point processes. 
    \item Introduces the hyperbolic function to model the rate at which relevant documents are found in a ranking. 
    \item Carries out experiments on a range of benchmark data sets to verify the effectiveness of the proposed approach and compare it against several alternative methods, including a generalised version of Cormack and Grossman's target method \cite{cormack2016engineering}. 
    \item Explores various configurations of the point process approach to discover the most effective. These configurations include two types of point process (Poisson Process and Cox Process) and four functions to model the rate at which relevant documents appear in a ranking (hyperbolic and three that have previously been used for this task: exponential function \cite{Sneyd19}, power law \cite{zobel1998reliable} and AP Prior distribution \cite{li2020stop}).
    \item Apply the proposed approach to a range of rankings, of varying effectiveness, to demonstrate its robustness. 
\end{itemize}

\section{Previous Work}\label{sec:background}

The problem of stopping methods for TAR has been discussed both within the literature associated with Information Retrieval and areas where document review tasks are commonly carried out such as eDiscovery \cite{yang2021minimizing,yang2021cost,lewis2016defining} and systematic reviewing, both in medicine \cite{shemilt2014pinpointing,wallace2013active} and other areas such as software engineering \cite{yu2019fast2} and environmental health \cite{Howard2020}.

Perhaps the most obvious approach to developing a stopping method is to estimate the total number of relevant documents in the collection, $\mathcal{R}$, and then stop when $\ell\mathcal{R}$ have been identified, where $\ell$ is the target recall. A number of approaches has been developed based on this strategy and are discussed in Section \ref{sec:no_rel}. Stopping methods that do not attempt to estimate $\mathcal{R}$ directly have also been described in the literature and we start by discussing these in Section \ref{sec:not_est}. 

\subsection{Stopping without Estimating $\mathcal{R}$}\label{sec:not_est}

The simplest stopping rules methods are based on heuristics such as stopping after a sequence of irrelevant documents has been observed, for example \citet{ros2017machine} stop after 50 consecutive irrelevant documents are observed. A range of similar approaches has been employed for the problem of deciding when to stop assessing documents during test collection development \cite{Losada2019}, including stopping after a fixed number of documents has been examined, stopping after a defined portion of all documents in the entire collection has been examined, stop after a fixed number of relevant (or non-relevant) documents has been observed and stop after a sequence of $n$ non-relevant documents has been observed. These approaches have the advantage of being straightforward to understand and implement.

The {\it knee method} \cite{cormack2016engineering} is designed to exploit the fact that relevant documents tend to occur more frequently early in the ranking and is based on the observation that examining additional documents often leads to diminishing returns. The approach makes use of a ``knee detection'' algorithm \cite{satopaa2011finding} to identify an inflection point in the gain curve produced by plotting the cumulative total of relevant documents identified against the rank. The slope ratio, $\rho$, at a point in the gain curve is computed as the gradient preceding that point divided by the curve gradient immediately following it. A suitable stopping point is one where the gradient drops quickly, i.e. a high slope ratio. \citet{cormack2016engineering} suggest 6 as a suitable value of $\rho$ in experiments where the target recall is 0.7. The effectiveness of the knee methods depends heavily on the value of $\rho$ used which may vary according to target recall and TAR problem; for example later work found that different values of $\rho$ were more effective \cite{li2020stop}.

\citet{di2018study} made use of the scores produced by a ranking algorithm (BM25) to predict the conditional probabilities of each examined document being relevant (or irrelevant) given the set of relevant (or irrelevant) documents identified so far. 
These values are then used to represent each document in a 2-dimensional space in which the stopping problem becomes one of finding a decision line in this space. 

A disadvantage of all these approaches is that they do not provide the reviewer with any information about the level of recall that has been achieved or confidence that that target recall has been achieved at the point that stopping is recommended. They may also rely on the values of key parameters (e.g. length of the sequence of irrelevant documents observed) and the most suitable values for these may vary between TAR problems. 

The {\it target method} \cite{cormack2016engineering} attempts to overcome these limitations with an approach that guarantees a target recall will be achieved with a specified confidence level. The approach proceeds by randomly sampling documents from the collection to identify a ``target set'' of relevant documents. Once these have been identified, all documents in the ranking are examined up to the final one in the target set. The number of relevant documents required for the target set is informed by statistical theory. \citet{cormack2016engineering} state that a target set size of 10 is sufficient to guarantee recall of 0.7 with 95\% confidence.\footnote{The original description of the target method \cite{cormack2016engineering} only provides details about the size of the target set required for these recall and confidence values. However, it is possible to generalise the argument for other levels of recall and confidence. Details are provided in Appendix \ref{appendix:target} (see also Section \ref{sec:baseline_target}).} The Quantile Binomial Coefficient Bound (QBCB) \cite{lewis2021certifying} approach is a variant of the target method which assumes that a control set of labelled documents is available. Like the target method, this approach specifies a minimum number of relevant documents that has to be identified from the control set before the method stops. This number is determined in a different way to the target method to avoid potential statistical bias from sequential testing. However, identifying a suitable control set can be a challenge in practice, particularly when prevalence is low as is often the case in TAR problems. 

A significant advantage of target and QBCB methods is the probability guarantees they provide about the target recall being achieved. However, one of their underlying assumptions is that the probability of a document being relevant does not vary through the ranking, which is unlikely to be the case in any reasonable ranking, leading to more documents than necessary being examined.

\subsection{Stopping by Estimating $\mathcal{R}$}\label{sec:no_rel}

An approach that has been explored by several researchers has been to examine documents up to a particular point in the ranking and then estimate the number of relevant documents remaining in some way, such as examining a sample (Section \ref{sec:sampling}), applying a classifier trained on the examined documents (Section \ref{sec:classification}) and using ranking scores (Section \ref{sec:score_dist}). Each approach is now discussed in turn. 

\subsubsection{Sampling Approaches}\label{sec:sampling}

Much of the work on sampling approaches for estimating $\mathcal{R}$ has been carried out within the context of work on systematic reviews in medicine. \citet{shemilt2014pinpointing} estimate the number of relevant documents remaining by sampling the unexamined ones. A statistical power size calculation was used to determine the size of the sample required in order to ensure that the estimate is within a desired level of confidence. Their approach was evaluated on two scoping reviews in public health, each of which involved the screening of extremely large sets of documents returned by queries ($> 800,000$). However, such an approach is sensitive to the estimate of the prevalence of relevant documents in the unexamined portion. 

\citet{Howard2020} describe a similar approach in which the number of relevant documents remaining is modelled using the Negative Binomial distribution. The number of relevant documents remaining is estimated simply as the total number of documents multiplied by the estimated probability of relevance, which is itself estimated by examining the documents most recently examined in the ranking. 
\citet{callaghan2020statistical} point out that the hypergeometric distribution is more appropriate for sampling without replacement and therefore better suited to model the situation that occurs when unexamined documents are sampled (since it would make no sense to return a document to the set of unexamined ones after a judgement on its relevance has been made). They combined the hypergeometric distribution with statistical hypothesis testing to develop a stopping rule that takes account of the desired confidence. Their approach was evaluated on a set of 20 systematic reviews from medicine and Computer Science that had been used in previous research on stopping criteria. 
 These approaches use established statistical theory to estimate the number of relevant documents remaining. However, they do not make use of the fact that, for any reasonable ranking, the probability of observing a relevant document decreases as the rank increases so they risk examining more documents than necessary. 
 
 The {\it S-CAL} \cite{cormack2016scalability} and {\it AutoStop} \cite{li2020stop} approaches address this by estimating $\mathcal{R}$ using nonuniform sampling strategies to reduce the number of documents that need to be examined. {\it S-CAL} \cite{cormack2016scalability} was developed within the context of a CAL system \cite{cormack2015autonomy} to produce an algorithm designed to achieve high recall for very large (potentially infinite) document collections. 
S-CAL examines a stratified sample across the collection where the inclusion probability decreases as the rank increases, rather than applying CAL to the entire collection. A classifier is used to carry out an initial ranking of the sample which is then split into batches. Relevance judgements are then obtained from a subset of documents within each batch and used to both estimate the number of relevant documents within that batch and as additional training data for the classifier. The algorithm proceeds until the number of relevant documents within each batch has been estimated and these figures are combined to estimate the total number of relevant documents. Similarly, {\it AutoStop} \cite{li2020stop} makes use of Horovitz-Thompson and Hansen-Huruwitz estimators \cite{horvitz1952generalization,thompson2012} 
 to provide unbiased estimates of $\mathcal{R}$ that take account of the decreasing probability of relevant documents being observed. (The Horovitz-Thompson estimator had been previously used to estimate the prevalence of relevant documents \cite{wallace2013active}, where it was shown to be more accurate than uniform random sampling, although that work did not go on to use the information provided to develop a stopping method.)  Stopping rules are based on either the estimator's direct output or this value with the variance added (to account for the estimate's uncertainty). The estimators employed by this approach rely on a suitable distribution for the sampling probabilities of each stratum of the sample, that is the probability of each document within that sample being relevant. \citet{li2020stop} found the AP-Prior distribution \cite{aslam2005measure,pavlu2007practical} as the best performing.

\subsubsection{Classification-based Approaches}
\label{sec:classification} 

As an alternative to sampling, which requires additional documents to be screened, recent approaches \cite{yu2019fast2,Yang2021heuristic} have used the relevance judgements from the observed documents as training data for a supervised classifier which is then used to estimate the number of relevant documents in the unobserved portion without the need for additional manual examination. These approaches are developed within ActiveLearning frameworks which already require the development of a classifier to rank unexamined documents so the extension to stopping rules represents limited additional effort. \citet{yu2019fast2} use the Support Vector Machine model employed within their Active Learning system to add ``temporary labels'' to the unexamined documents which are used to train a logistic regression classifier used to estimate the total number of relevant documents in the unexamined portion. \citet{Yang2021heuristic} present a similar approach in which a logistic regression classifier is trained on the observed documents and applied to the unobserved potion. A point estimate of the total number of relevant documents is calculated together with an estimate of its variance and used to produce two stopping rules: one based on the point estimate of the total number of documents and another where twice the variance of this estimate is added (equating to approximately a 95\% confidence interval on the estimate). 
This method is essentially an example of the ``classify and count'' approach to the more general problem of volume estimation. However, \citet{del2021learning} pointed out that this approach is sub-optimal, not least because the prevalence of relevant documents in the observed and unobserved portions are likely to differ. 

\subsubsection{Score Distribution Approaches}\label{sec:score_dist}

\citet{hollmann2017ranking} made use of the scores assigned by a ranking algorithm to estimate $\mathcal{R}$. Following a standard approach \cite{arampatzis2009stop}, the distribution of relevant documents is modelled as a Gaussian random variable and used to compute the probability of each document being relevant based on the score assigned to it by the ranking algorithm. The total number of relevant documents at each point in the ranking can then be estimated by summing these probabilities and this information is used to identify when a particular level of recall has been achieved. \citet{cormack2009machine} fitted a normal distribution to the scores of the relevant documents that had been identified and used the area under the curve to estimate $\mathcal{R}$. These approaches, and \citet{di2018study} (see Section \ref{sec:not_est}), are applications of score distribution methods \cite{arampatzis2009stop,kanoulas2010score} to the stopping problem. 

\subsection{Summary} 

Some approaches to the TAR stopping problem are based on simple heuristics that may be effective under certain circumstances but are not likely to be generally reliable (see Section \ref{sec:not_est}). Attempts have been made to develop more robust stopping rules that offer some assurance that the target recall has been reached. The most straightforward way of achieving this is to estimate the total number of relevant documents but this generally proves to be expensive with large numbers of documents having to be examined to achieve the levels of statistical reliability that are sought (see Section \ref{sec:no_rel}). Current methods that provide assurance without estimating $\mathcal{R}$ directly (e.g. target method \cite{cormack2016engineering} and QBCB \cite{lewis2021certifying}) do not model the fact that the prevalence of relevant documents is likely to reduce substantially with the ranking which also leads to more documents being examined than necessary. 

This paper provides an alternative approach to the stopping problem that makes use of a well-established stochastic model (counting processes) to estimate $\mathcal{R}$ and thereby produce a stopping criterion. The approach has the advantage that the estimate can be made by examining the top ranked documents, which are most likely to be relevant, thereby reducing the overall number of documents that need to be examined.

\section{Point Processes}\label{sec:point_processes}

In their most general sense, point processes can be viewed as a stochastic model of a random element (i.e. generalised random variable) defined over a mathematical space and with values that can be considered as ``points'' within that space \cite{cox1980point,taylor1994}. They are often applied within spatial data analysis where they have been applied to a wide range of disciplines including epidemiology, seismology, astronomy, geography and economics. Point processes defined over the positive integers have proved to be particularly useful since they can be used to model the occurrences of random events in time, for example, the arrival of customers in a queue, emissions of radioactive particles from a source or impulses from a neuron. Applications within Computer Science include queuing theory \cite{bhat2015introduction}, computational neuroscience \cite{miller2018introductory}, social media analytics \cite{lukasik2015point} and modelling user interaction with recommendation systems \cite{thomas2018bayesian}.

In the application of point processes described here, the space is a ranking of documents and the random event is the occurrence of a relevant document in this ranking. The description of point processes which follows, therefore focuses on this application rather than considering more general types of point processes. 

We begin by introducing the point processes used in this paper (Sections \ref{sec:point_processes} and \ref{sec:cox}) and then describe candidate models for the occurrence of relevant documents (Section \ref{sec:rate_function}).

\subsection{Poisson Processes}\label{sec:point_processes}

Poisson Processes \cite{Poisson} are an important type of point process which assume that events occur independently of one another and the number of occurrences in a given interval follows a Poisson distribution. They are suitable for situations that can be modelled as a large number of Bernoulli trials with a low probability of success in each trial \cite{taylor1994}, such as  TAR problems where the prevalence of relevant documents is normally very low. Poisson Processes can be used to estimate the number of occurrences of relevant documents found within some portion of the ranking. The average frequency with which relevant documents are observed is denoted by a parameter $\lambda$ which is referred to as the rate and assumed to be greater than 0.

\subsubsection{Homogeneous Poisson Processes} The simplest type of Poisson Process, a {\it homogeneous Poisson processes}, is produced when the value of $\lambda$ is constant. Then, the number of relevant documents that has occurred at point $t$ in the ranking, $N(t)$, is modelled by a Poisson distribution with the parameter $\lambda t$, i.e. the probability that $n$ relevant documents have been observed by rank $t$ is given by 
\begin{equation}\label{ip_proc}
	   P\left(N(t) = n\right) = \frac{\left[ \lambda t \right]^n}{n!} e^{-\lambda t}.
\end{equation}

In addition, $N(i, j)$, the number of relevant documents between ranks $i$ and $j$, is a Poisson distribution with the parameter $\lambda (j - i)$ with the probability that number equals $n$ given by 
\begin{equation}\label{ip_proc_2}
	   P\left(N(i, j) = n\right) = \frac{\left[ \lambda (j - i) \right]^n}{n!} e^{-\lambda (j - i)}.
\end{equation}

Consider a simple illustrative example where we wish to estimate the number of relevant documents between ranks 10 and 100 where $\lambda = 0.05$. Then 

\begin{equation}
P(N(10, 100) = n) = \frac{\left(0.05 \times 90\right)^n}{n!} e^{-0.05 \times 90} = \frac{4.5^n}{n!} e^{-4.5}\;, 
\end{equation} 
i.e. a Poisson distribution with a mean of 4.5.

\subsubsection{Inhomogeneous Poisson Processes}\label{sec:ipp}

Assuming that the rate at which relevant documents are observed stays constant is not reasonable in practice since for any reasonable retrieval system relevant documents are more likely to be found earlier in the ranking. This can be taken account of using a rate that varies as a function of the ranking to produce an {\it Inhomogeneous Poisson Process}. Let $\lambda(x)$ be a rate function where $x$ is a position in a ranking, i.e. $x \in \{1, 2, 3, \ldots, N \}$ for a ranking of $N$ documents.\footnote{Candidate rate functions are discussed in Section \ref{sec:rate_function}.}
$\Lambda(a,b)$ is defined as the integral of the rate function, $\lambda(x)$, between ranks $i$ and $j$, i.e.
\begin{equation}\label{eq:eqnLam}
\Lambda(i, j) = \int_i^j \lambda(x)dx.
\end{equation}
Then $N(t)$ is a modelled as a Poisson distribution with the parameter $\Lambda(0, t)$, that is the probability of $N(t)$ having the value $n$ is given by: 
\begin{equation}\label{eq:ip_proc_all}
	   P\left(N(t) = n\right) = \frac{[\Lambda(0,t)]^n}{n!} e^{-\Lambda(0,t)}.
\end{equation}

In addition, the number of relevant documents between ranks $i$ and $j$, $N(i, j)$, is a Poisson random variable with parameter $\Lambda(i, j)$, so the probability of observing $n$ relevant documents is given by: 
\begin{equation}\label{eq:ip_proc_range}
	   P\left(N(i, j) = n\right) = \frac{[\Lambda(i, j)]^n}{n!} e^{-\Lambda(i, j)}.
\end{equation}

For example, if the rate function is $\lambda(x) = x^{-2}$ and we, again, wish to estimate the number of relevant documents between ranks 10 and 100. Then $\Lambda(10, 100) = 0.09$ so $P(N(10, 100) = n) \sim Poisson(0.09)$. 

\subsection{Cox Processes}\label{sec:cox}

In our application, the rate function, $\lambda(x)$, represents the probability of a relevant document being observed at a particular rank, which is not straightforward to estimate. Cox processes \cite{cox1955some}, also known as doubly stochastic Poisson processes, are an extension of Poisson Processes that take account of uncertainty about the rate function. Rather than being a fixed function, as in a Poisson Process, the rate function in a Cox Process is modelled as a probability distribution over possible rate functions, $P(\lambda)$. The random variable representing the number of relevant documents that occur between ranks $i$ and $j$ is then defined by computing the expected value of \ref{eq:ip_proc_range} given $P(\lambda)$, i.e. 
\begin{equation}\label{eq:cox_proc}
P\left(N(i, j) = n\right) = \int_0^{\infty} \frac{[\Lambda(i, j)]^n}{n!} e^{-\Lambda(i, j)} P(\lambda) \;d\lambda
\end{equation}
where $P(\lambda)$ is the probability of the rate function taking a particular value so that $P\left(N(i, j) = n\right)$ is estimated by integrating over all possible values of $\lambda$. 

In practice, a general form is chosen for the rate function, for example, $\lambda(x) = x^{-a}$ where $a$ is a parameter used to select particular functions. So, $a = 2$ would give the function $\lambda(x) = x^{-2}$ used in the Inhomogenous Poisson Process example (Section \ref{sec:ipp}). 
The parameters of the rate function are assigned values from some probability distribution, which produces a distribution over possible rate functions, $P(\lambda)$.

\subsection{Rate Functions}\label{sec:rate_function} 

Selecting an appropriate general form for the rate function 
is a key decision in the application of point processes. 
An appropriate function should assume that a suitable ranking has, in accordance with the probability ranking principle \cite{robertson1977probability}, succeeded in placing documents that are more likely to be relevant higher in the ranking than those less likely to be and, consequently, the rate at which relevant documents occur decreases in direct proportion to the document's position in the ranking. A range of suitable functions exist which we now discuss. 

\subsubsection{Exponential function}\label{sec:exponential}
The mathematical properties of the exponential function make it a convenient choice of rate function.
It is defined as 
\begin{equation}\label{eq:exponential}
\lambda(x) = ae^{bx},
\end{equation}
where $x$ is an index in a ranking (i.e. $x \in \{1, 2 \ldots N \}$ for a collection of $N$ documents) and $a, b \in \mathbb{R}$ are parameters controlling the function's shape. Substituting into Equation~\ref{eq:eqnLam}, the expected number of relevant documents between index $i$ and index $j$ is given by: 
\begin{equation}\label{eq:LamExp}
\Lambda(i, j)  = \int_i^j ae^{bx} dx \\  = \frac{a}{b}\left(e^{bj} - e^{bi} \right).
\end{equation}

Combining Equations~\ref{eq:ip_proc_range} and \ref{eq:LamExp}, the probability of observing $n$ relevant documents between ranks $i$ and $j$ is given by: 
\begin{equation}\label{eq:ip_combined}
	   P\left(N(i, j) = n\right) = 
	   \frac{\left[\frac{a}{b}\left(e^{bj} - e^{bi}\right)\right]^n}{n!} e^{-\left(\frac{a}{b}\left(e^{bj} - e^{bi}\right)\right)}.
\end{equation}

Equation~\ref{eq:ip_combined} provides a convenient and easily computable closed form solution for estimating the number of relevant documents.

\subsubsection{Hyperbolic Decline} 

The hyperbolic decline function also meets the criteria for a suitable rate function. It is widely used in the field of petroleum engineering to model declining productivity of oil and gas wells to predict future output \cite{arps1945analysis} but, to the best of our knowledge, has not previously been used in IR. The function is defined as:
\begin{equation}\label{eq:hyperbolic}
\lambda(x) = \frac{a}{(1 + bcx)^{\frac{1}{b}}}
\end{equation}
where $x$ is, once again, an index in a ranking and $a$, $b$ and $c$ are parameters controlling the shape of the function with $0 \leq b \leq 1$. Note that when $b = 0$ equation \ref{eq:hyperbolic} becomes equivalent to exponential decline (Section \ref{sec:exponential}) while $b = 1$ produces a harmonic decline function. 

Integrating equation \ref{eq:hyperbolic} produces: 
\begin{equation}\label{eq:hyperbolic_int}
\int_i^j \frac{a}{(1 + bcx)^{\frac{1}{b}}} dx = 
\left[ \frac{a}{c(b - 1)}\left( 1 + bcx \right)^{1 - \frac{1}{b}} \right]_{i}^{j} = 
\frac{a}{c(b - 1)} \left( \left( 1 + bcj \right)^{1 - \frac{1}{b}} - \left( 1 + bci \right)^{1 - \frac{1}{b}} \right)
\end{equation}

Equation \ref{eq:hyperbolic_int} can be substituted into equation~\ref{eq:ip_proc_range} in a similar way to the exponential function (see Section \ref{sec:exponential}) to create a random variable to estimate the number of relevant documents in a portion of the ranking.

\subsubsection{Power Law} Power laws have been proposed as a suitable model of the rate at which relevant documents are observed in a ranking \cite{zobel1998reliable} and have been shown to be useful for estimating the number of relevant documents remaining for test collection development, e.g. \cite{Losada2019}. 
Power laws have the form 
\begin{equation}
   \lambda(x) = ax^{b} 
\end{equation} where $x$ is an index in the ranking and the parameters $a, b \in \mathbb{R}$ determine the function's shape. Substituting this into equation \ref{eq:eqnLam} produces
\begin{equation}\label{power_integral}  
\Lambda(i, j) = 
	    \int_a^b ax^{b} dx  = \left[ \frac{ax^{b+1}}{b + 1}\right]_{i}^{j} =
	    \frac{a}{b + 1}\left(j^{b+1} - i^{b+1}\right)
\end{equation}

which can also be substituted into Equations~\ref{eq:ip_proc_range} and \ref{eq:cox_proc} in a similar way to the previous rate functions.

\subsubsection{AP Prior Distribution} The AP-Prior distribution \cite{aslam2005measure,pavlu2007practical} has been applied in IR evaluation and demonstrated to be a suitable prior for the relevance of documents in a ranked list \cite{aslam2007query,aslam2003unified,aslam2006statistical,li2017active,yilmaz2006estimating,yilmaz2008simple}. It was also used in the AutoStop algorithm \cite{li2020stop} (see Section \ref{sec:no_rel}). The AP-Prior distribution models the probability of relevance at each rank based on its contribution to the average precision score 
\begin{equation}\label{eq:apprior}
 \lambda(x) = \frac{1}{Z}\log\frac{n}{x}, \hspace{2em} Z = \sum_{i=1}^n\log{\frac{n}{x}} 
\end{equation}
where $x$ is (again) an index in the ranking, $n$ the total number of documents in the collection and $Z$ a normalisation factor. The integral of Equation \ref{eq:apprior} is easier to derive after some rearrangement:

\begin{equation}
\lambda(x) = \frac{\log\frac{n}{x}}{\sum_{i=1}^n\log{\frac{n}{x}}}  = \frac{\log\frac{n}{x}}{\sum_{i=1}^n\left( \log{n} - \log{x} \right) }
 = \frac{\log\frac{n}{x}}{ \left( n.\log{n} - \log{n!} \right) } \;. 
\end{equation}

So,
\begin{equation}\label{eq:apprior_derivation}
\Lambda(i , j) = 
\int_i^j \frac{\log\frac{n}{x}}{n.\log{n} - \log{n!} } dx = 
\left[ \frac{x \log{\frac{n}{x}}}{n.\log{n} - \log{n!}} \right]_i^{j} = 
\frac{j \log{\frac{n}{j}} - i \log{\frac{n}{i}}}{n.\log{n} - \log{n!}} \;.
\end{equation}

Unlike the other rate functions, the AP-Prior is a probability distribution, i.e. sums to 1 over all documents in the ranking. To provide a point process rate function it needs to be scaled based on the expected total number of relevant documents in the ranking which can be achieved by multiplying Equation~\ref{eq:apprior_derivation} by a scalar, $a$. The value of $a$ then becomes a parameter controlling the function's shape similar to the parameters that control the shape of the exponential, hyperbolic and power law rate functions. 
This rate function can then be combined with Equations~\ref{eq:ip_proc_range} and \ref{eq:cox_proc} to produce a point process using the same approach that was used for the other rate functions.

\section{Stopping Algorithm}\label{sec:algorithm} 

The point process framework described in the previous section allows us to define a stopping method. Briefly, the approach operates by screening the top ranked documents, from rank 1 to rank $k$, and counting the number of relevant documents, referred to as $rel(1, k)$. The point process is then used to estimate the number of relevant documents in the remaining (i.e. unexamined) part of the ranking (i.e. ranks $k+1$ $\ldots$ $n$, where $n$ is the total number of documents), denoted as $\widehat{rel}(k+1, n)$. An estimate of the total number of relevant documents in the entire ranking, $\mathcal{\hat{R}}$, is then given by $\mathcal{\hat{R}} = rel(1, k) + \widehat{rel}(k+1, n)$ and this value used to estimate the number of relevant documents required to reach a given target recall, $\ell$, i.e. $ \ell \mathcal{\hat{R}}$. The algorithm stops if a sufficient number of relevant documents has been found to reach the desired level of recall (i.e. $rel(1, k) \geq \lceil \ell \mathcal{\hat{R}} \rceil$, where $\lceil . \rceil$ is the ceiling function), otherwise the process is repeated after screening more of the documents (i.e. increasing the value of $k$). 

\begin{figure}[!ht]
\centering
\includegraphics[width=0.8\textwidth]{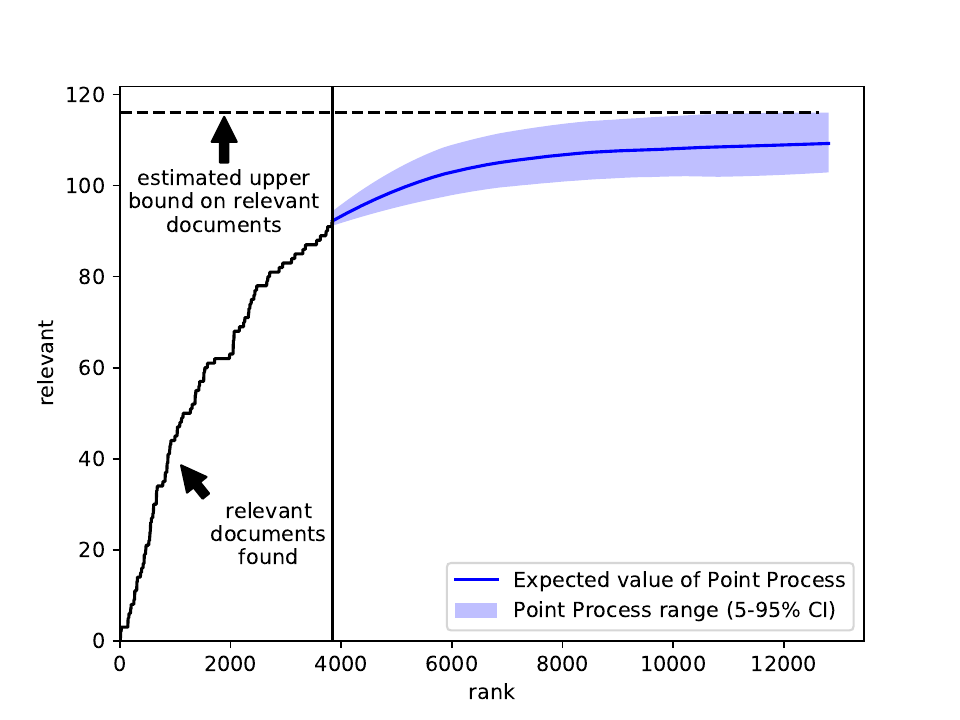}
\caption{Representation of a Point Process applied to a ranked set of 12,807 documents of which 114 are relevant. The figure is divided into two parts by the vertical line just below rank 4000. Documents to the left of this line have been screened for relevance and the figure shows the cumulative number of relevant documents identified at each point of the ranking. Documents to the right of the line have not yet been examined and the figure illustrates a Point Process used to estimate the number of relevant documents. The shaded area represents the number of relevant documents predicted by the Poisson Process in the 5\% to 95\% confidence range. Taking the upper bound of this estimate for the final document in the ranking produces a prediction of the total number of relevant documents in the collection.}
\label{fig:pp_example}
\end{figure}

A key part of this process is using the point process to estimate $\mathcal{R}$. Documents that have been examined are analysed to estimate the probability of a relevant document being encountered at each point in the ranking and one of the rate functions described in Section \ref{sec:rate_function} fitted (see Section \ref{sec:rate} for additional details about this process). This rate function is then used to produce a point process that estimates the number of relevant documents that will be encountered by any point in the unexamined documents found later in the ranking by modelling this value as a Poisson random variable. If the first $k$ documents in a ranking of $n$ documents have been screened then the number of relevant documents in the unscreened portion of the ranking is modelled as the random variable $N(k+1, n) \sim Poisson()$. By examining its cumulative distribution function (CDF), it is possible to estimate the maximum value of $N(k+1, n)$ with some desired level of probability, $p$. 

A visualisation of the process is shown in Figure \ref{fig:pp_example}.

\subsection{Fitting the Rate Function}\label{sec:rate}

A set of points representing estimates of the probability of encountering a relevant document in the screened documents (i.e. ranked from 1 to $k$) are created by averaging the number of relevant documents observed within a sliding window. Non-linear least squares \cite{kelley1999iterative} is then used to find the parameters of the rate function being used that best fit that data. 

While the rate function is fitted using information derived from the first $k$ documents, it is then extrapolated across the entire ranking using the parameters produced by the fitting process and the estimates of the probability of observing a relevant document it produces used by the point process. Since it is important to ensure that these estimates are reliable, the rate function is only fitted if a sufficient number of relevant documents has already been observed in the first $k$ documents. Two different approaches to determining this were applied. Firstly, the ``static'' approach checked if a fixed number of relevant documents had been observed. The values 10 and 20 were explored, so the rate function would only be fitted if $rel(1, k) \geq 10$ and $rel(1, k) \geq 20$, respectively. The second, ``dynamic'' approach reduced the number of relevant documents that had to be observed as the ranking increased, so that the rate function would only be fitted if $rel(1, k) \geq 20 \times (1 - \frac{k}{n})$, where $n$ is the total number of documents in the ranking. 

If enough relevant documents have been observed and a rate function fitted then an additional check is carried out by measuring the difference between the observed values and those predicted by the rate function using Normalised Root Mean Squared Error (NRMSE). NRMSE measures the difference between observed values and those predicted by a model by computing the average of the squared differences between them and then normalising that value by the range of the observed data. More formally, NRMSE is computed as 
\begin{equation}\label{eq:nrmse}
\frac{\sum_{i} \left( \hat{y}_{i} - y_{i} \right)^{2}}{y_{max} - y_{min}}   
\end{equation}
where $y_{i}$ and $\hat{y}_{i}$ are (respectively) the observed and predicted values for the probability of document relevance while $y_{max}$ and $y_{min}$ are (respectively) the highest and lowest observed probabilities of document relevance. Note that NRMSE is computed using only the observed and predicted values for the first $k$ documents (i.e. those which have already been screened). If the NRMSE value exceeds a threshold, then the fitted curve is not considered to be an accurate model for the true rate. If this happens the algorithm does not attempt to compute the point process, since its results may not be reliable, and there is no attempt to estimate the total number of relevant documents until further documents have been screened. Several values for this threshold are explored in the experiments reported later (see Section \ref{sec:hyperparameter}). 

If the point process being used is an inhomogeneous Poisson Process (Section \ref{sec:ipp}) then the point estimates of the rate function parameters can simply be supplied to the closed form of the relevant integral shown in Section \ref{sec:rate_function}. However, when a Cox Process (Section \ref{sec:cox}) is being used, the point process also considers the estimated variance of these parameters. The parameters of the rate function are modelled using a normal distribution, $\mathcal{N}( \mu, \sigma)$, where $\mu$ is the least square estimate of the parameter value and $\sigma$ its estimate of the variance. Unfortunately, no convenient closed forms exist for the integrals required to compute equation~\ref{eq:cox_proc} and instead, they are computed using numerical integration (Simpson's rule). 

\subsection{Stopping Algorithm Pseudocode}\label{sec:pseudo}

\begin{algorithm}[h]
\caption{Algorithm to Identify Stopping Rank}\label{stopping_algo}
\begin{algorithmic}[1]
\STATE {\bf Input:} n (= no. documents in ranking), $\ell$ (= target recall level, e.g. 0.7), p (= confidence level, e.g. 0.95), $\alpha$ (= initial sample size, e.g. 0.05), $\beta$ (= sample increment size, e.g. 0.025)\label{alg:input}
\STATE {\bf Output:} k (= stopping rank)\label{alg:output}
\STATE $k \gets \alpha \times n $ \label{alg:init}
\STATE Obtain relevance judgements for ranks 1 to $k$  \label{alg:rel_j} \COMMENT{Note: Some may have been obtained during previous iterations}
\WHILE{$k < n$}
\STATE $rel(1, k) \gets$ count of relevant documents found in ranks 1 to $k$\label{alg:rel_count}
\IF {$rel(1, s) \geq threshold$} \label{alg:rel_check}
\STATE Fit rate function to documents in ranks 1 to $k$ \label{alg:fit_rate}
\IF {NRMSE of fit $\leq threshold$} \label{ref:nrmse}
\STATE Apply point process to documents ranked $k+1$ to $n$ \label{alg:pp}
\STATE $\widehat{rel}(k+ 1, n) \gets$ estimate of relevant documents between ranks $k+1$ to $n$ \label{alg:est_r} 
\STATE $\mathcal{\hat{R}} \gets rel(1, k) + \widehat{rel}(k+ 1, n)$ \label{alg:r_total}
\IF {$  \ell \mathcal{\hat{R}} < rel(1, k) $} \label{alg:test_r}
\STATE goto line \ref{alg:return} \label{alg:break}
\ENDIF
\ENDIF
\ENDIF
\STATE Obtain relevance judgements for ranks $k + 1$ to  $\left( k + 1 \right) + \left( \beta \times n \right)$  \label{alg:rel_j2}
\STATE $k \gets k + \left( \beta \times n \right)$ \label{alg:iterate}
\ENDWHILE
\STATE return $k$ \label{alg:return}
\end{algorithmic}
\end{algorithm}

Pseudocode for the stopping method is shown in Algorithm \ref{stopping_algo}. The method is provided with several pieces of information (line \ref{alg:input}). The target recall level ($\ell$) and confidence levels ($p$) indicate the desired level of recall and the algorithm's confidence that this has been achieved prior to stopping. The total number of documents in the ranking ($n$) must also be provided together with parameters controlling the number of documents that are examined between each application of the point process to check whether the stopping point has been reached ($\alpha$ and $\beta$). The values of $\alpha$ and $\beta$ could be adjusted to check whether the stopping point has been reached as frequently as required, and potentially for every document in the ranking, although more frequent estimates would increase the computational cost. The algorithm outputs a rank at which it estimates that the recall and confidence ($\ell$ and $p$) targets have been met so the screening of documents can cease (line \ref{alg:output}).

The algorithm begins by obtaining relevance judgements for the top ranked documents (lines \ref{alg:init} and \ref{alg:rel_j}) and counting the number which are relevant (line \ref{alg:rel_count}). At this point there is a check whether enough relevant documents have been found to attempt to fit a rate function (line \ref{alg:rel_check}), if not then the number of screened documents is gradually increased until enough have been found. Assuming that a sufficient number of relevant documents has been found, a rate function is fitted (line \ref{alg:fit_rate})
and checked by computing the NRMSE (line \ref{ref:nrmse}) (see above). A point process is run and its output used to estimate the number of relevant documents in the portion of the ranking that has yet to be screened (lines \ref{alg:pp} and 
\ref{alg:est_r}). This information is then used to estimate the total number of relevant documents in the entire ranking (line \ref{alg:r_total}). The algorithm stops and returns the current rank if enough relevant documents have already been observed to achieve the target recall (line \ref{alg:break}), otherwise the next highest ranking documents in the ranking are screened and the process is repeated (lines \ref{alg:rel_j2} and \ref{alg:iterate}). The process of increasing the number of screened documents continues until either the algorithm concludes that the target recall has been reached or all documents have been screened.

\subsection{Properties of Approach} 

The stopping method outlined above has a number of advantages. Firstly, the screening effort is focused on the top ranked documents, in other words, those which are most likely to be relevant. Unlike some other approaches (e.g. \cite{cormack2016engineering,li2020stop}), there is no need to obtain relevance judgements for documents sampled from across the ranking, thereby leading to additional effort that is unlikely to identify additional relevant documents. Secondly, the proposed method provides an estimate of the number of relevant documents at any point in the ranking using a well understood statistical model. 
Consequently, recall can be estimated for any point of the ranking, unlike approaches that identify a stopping point only for a pre-specified ranking (e.g. \cite{cormack2016engineering}) or provide an estimate of the recall achieved at a particular rank (possibly with an associated confidence value) but do not provide information about the recall that is likely to be achieved after further documents have been screened (e.g. \cite{shemilt2014pinpointing,Yang2021heuristic}). Finally, the computational effort required is relatively modest. The stopping point is identified by examining the distribution of the estimated number of relevant documents which can be calculated from the expression produced by combining one of the rate functions with the point process (see, for example, Equation~\ref{eq:ip_combined}).

\section{Evaluation}

\subsection{Baselines}\label{sec:baselines} 

Comparison of stopping algorithms has been a challenging problem since they are deployed within different retrieval frameworks, each of which uses its own ranking, and using different implementations. The situation has improved recently with the release of reference implementations for a range of stopping methods \cite{li2020stop,yang2022tarexp}. Although we found that integrating new approaches into these frameworks was less straightforward than we had hoped, we were able to extract the rankings used by the reference implementations described by \citet{li2020stop} which allowed us to directly compare the performance of our approach against a range of alternative approaches, particularly \citet{cormack2016engineering} and \citet{li2020stop} (see Section \ref{sec:rankings}).

We compared the proposed method against several previous approaches described in Section \ref{sec:background} with results produced using the reference implementation provided by \citet{li2020stop}: {\bf target}\footnote{\citet{li2020stop} chose to treat the size of the target set as a hyper-parameter which was set to 5 based on tuning on training data although setting the target set in this way does not provide the theoretical performance guarantees that were one of the motivations behind this method \cite{cormack2016engineering}.} \cite{cormack2016engineering}, {\bf knee} \cite{cormack2016engineering}, {\bf SCAL} \cite{cormack2016scalability}, {\bf AutoStop} \cite{li2020stop}, {\bf SD-training} and {\bf SD-sampling} \cite{hollmann2017ranking}.\footnote{SD-training and SD-sampling are variants of the score distribution method described by \citet{hollmann2017ranking} which vary in how they identify the relevant documents required to model ranking scores. The first variant uses training topics while the second samples documents to obtain relevance judgements from the user.} Results for the {\bf QBCB} method \cite{lewis2021certifying} are also reported. We were unable to find a reference implementation for this approach so results were produced using a modified version of the \citet{li2020stop} implementation of the target method. The QBCB method requires the size of the control set to be specified. We chose a value of 50 to balance accuracy and cost with the relevant documents included in the set identified by random sampling. The QBCB method assumes that the control set is provided to the algorithm but this can be challenging for high target recalls, for example for a target recall of 0.9 the control set would need to contain 49 relevant documents. We obtain these by randomly sampling until the required number of relevant documents for the control set has been identified but do not include any sampled documents that occur after the algorithm stops in the calculations of the algorithm's cost. This optimistic assumption about the availability of a control set benefits the QBCB algorithm.

\subsubsection{Adapted target method ({\bf TM-adapted})}\label{sec:baseline_target} We also experimented with a more generalised version of the target method. The original description of this approach shows that a target size of 10 relevant documents is sufficient to achieve recall $\geq$ 0.7 with 95\% confidence \cite{cormack2016engineering} but does not state the number that needs to be identified for other recall and confidence levels. By generalising the argument in \citet{cormack2016engineering}, it can be shown that the required number is 
\begin{displaymath} 
- \frac{\log(1 -c)}{1 - \ell}
\end{displaymath}

where $\ell$ is the desired level of recall (e.g. 0.7) and $c$ is the confidence in this level of recall being achieved (e.g. 0.95). (See Appendix \ref{appendix:target} for details about how this result was derived.) For example, 30 relevant documents must be observed in the random sample when $\ell$ = 0.9 and $c$ = 0.95. Results for this approach are generated by varying the target number using the same reference implementation used for the standard target method \cite{li2020stop}. 

\subsubsection{{\bf Oracle}} Results of an Oracle approach are also reported. The oracle starts at the top ranked document and continues through the ranking until enough documents have been observed to reach the desired recall. This approach is not practically feasible since it assumes complete information about the ranking (including the relevance of documents beyond those that have been observed). However, it is useful to provide context for other methods by indicating the minimum number of documents that need to be examined in a fixed ranking in order to achieve the desired recall. This number will vary according to the individual ranking - it will be lower when relevant documents have been ranked highly and lower when they have not - and places a limit on how early a method can stop while achieving the target recall. 

It is worth noting that the recall achieved by the oracle method can be higher than the target recall under certain circumstances. This happens when the number of documents in the topic makes it impossible to stop at the target recall exactly and in these cases the oracle stops at the lowest possible recall above the target. For example, if the target recall is 0.8 and a topic contains 11 relevant documents then the oracle method will stop after 9 relevant documents have been identified, representing a recall of approximately 0.818.

\subsection{Evaluation Metrics}\label{sec:metrics}

Stopping methods aim to identify a set portion (possibly all) of the relevant documents in a collection, the target recall, while requiring that as few documents as possible are manually examined. This can be viewed as a multi-objective optimisation problem that aims to both (1) maximise the probability that the number of documents identified is at least the target recall, and (2) minimise the number of documents that need to be manually reviewed. These objectives are generally in opposition since increasing the probability of achieving the target recall normally requires more documents to be reviewed, and vice versa, making it difficult to summarise them using a single metric. 

A wide range of metrics has been used to evaluate previous work on stopping methods. 
To simplify comparison with previous work, we adopt the same metrics as those reported previously \cite{li2020stop}. All metrics were computed using the {\tt tar\_eval.py} script provided for the CLEF Technology Assisted Review in Empirical Medicine tasks.\footnote{Available from \url{https://github.com/CLEF-TAR}}

{\bf Recall}: The recall metric is the proportion of relevant documents identified. It is defined as: 
\begin{equation}
    recall = \frac{r}{\mathcal{R}}
\end{equation}

where $r$ is the number of relevant documents identified and $\mathcal{R}$ the total number of relevant documents in a collection. 

{\bf Cost}: The cost metric is the proportion of the total documents in the collection that have to be manually reviewed before a stopping point is identified. It is defined as: 
\begin{equation}
    cost = \frac{o}{n}
\end{equation}

where $o$ is the number of documents that need to be examined and $n$ is the total number of documents in the collection. 

{\bf Reliability}: The reliability metric, due to \citet{cormack2016engineering}, is the proportion of topics in a collection where an approach achieves the target recall. Let $\mathcal{C}$ be a collection of topics, then the reliability of an approach over $\mathcal{C}$ is given by: 

\begin{equation}
    reliability = \frac{| \{ c \in \; \mathcal{C} : recall \geq recall_{t} \} |}{|\mathcal{C}|}
\end{equation}

where $recall_{t}$ is the target recall. The reliability metric is unique among those used in this work in that it is defined over a collection of topics, rather than a single topic. 

{\bf Relative error}: The relative error metric is the normalised absolute difference between the recall achieved by a stopping method and the target recall. It is defined as 

\begin{equation}
    Relative\;error\;(RE) = \frac{|recall - recall_{t} | }{recall_{t}}.
\end{equation}

$\mathbf{loss_{er}}$:
The $loss_{er}$ metric \cite{cormack2016engineering} is designed to be a single metric that captures the two objectives for the stopping task. Its development was informed by experience from TREC Total Recall tracks \cite{roegiest2015trec,grossman2016trec} and also adopted by the CLEF Technology Assisted Reviews in Empirical Medicine Tracks \cite{kanoulas2017clef,kanoulas2018clef,kanoulas2019clef}. The $loss_{er}$ measure is the sum of two components: $loss_{r}$ and $loss_{e}$. The first of these is defined as a quadratic loss function that penalises a method for failing to achieve 100\% recall: 
\begin{equation}
    loss_{r} = \left( 1 - recall \right)^{2}.
\end{equation}

It is worth mentioning that $loss_{r}$ assumes that a method aims to achieve 100\% recall. While this might be desirable in many circumstances, it might not always be the case. It would be straightforward to adapt $loss_{r}$ to only penalise a method only when its recall is below a set target recall (i.e. $recall_{t}$), but we choose not to adjust the method to simplify comparison with previous work and because the relative error measure already captures information about the difference between the achieved and target recall. The second component, $loss_{e}$, motivated by experience from the TREC 2015 Total Recall Track \cite{roegiest2015trec}, is defined as: 
\begin{equation}
    loss_{e} = \left( \frac{100}{n} \right)^{2} \left( \frac{o}{R + 100} \right)^{2}.
\end{equation}

This metric is motivated by the observation that a ``reasonable'' effort might be given by $aR + b$ where $a$ represents an effort proportional to the total number of relevant documents and $b$ a fixed cost. (Note that $a = 1$ and $b = 0$ is the ideal scenario where effort is minimised as far as possible.) The values of $a$ and $b$ are somewhat arbitrary, previous work \cite{cormack2016engineering} suggested that $a \leq 2$ and $b \leq 1000$ would be a reasonable effort to achieve recall $\geq$ 0.7 with 95\% confidence. We follow the CLEF Technology Assisted Reviews in Empirical Medicine Tracks \cite{kanoulas2017clef,kanoulas2018clef,kanoulas2019clef} and \citet{li2020stop} in choosing $a = 1$ and $b = 100$. Then the $\frac{o}{R + 100}$ element represents the proportion of documents examined compared to a ``reasonable'' effort. The $\frac{100}{n}$ element is a weight that determines the importance of this type of loss. (See \citet{cormack2016engineering} for further discussion of the motivation behind $loss_{er}$.)

The $loss_{er}$ measure itself is defined as the sum of the $loss_{e}$ and $loss_{r}$ components: $loss_{er} = loss_{e} + loss_{r}$.

\subsection{Datasets}\label{sec:datasets}

Evaluation is carried out using common benchmark data sets representing TAR problems from a range of domains: the CLEF Technology-Assisted Review in Empirical Medicine, the TREC Total Recall Tasks and the TREC Legal Tasks. The data sets used are the same as those used in previous work \cite{li2020stop} to facilitate comparison. 

{\bf CLEF Technology-Assisted Review in Empirical Medicine}\footnote{Available from \url{https://github.com/CLEF-TAR}} The CLEF task on TAR in empirical medicine focused on the identification of evidence for systematic reviews. These reviews support evidence-based approaches to medicine by identifying, appraising and synthesising and summarising current knowledge in relation to a research question, for example {\it Rapid diagnostic tests for diagnosing uncomplicated P. falciparum malaria in endemic countries} \cite{abba2011rapid}. Identification of as much relevant evidence as possible is a key priority in systematic review development. 

The task was run from 2017 to 2019 and three data sets were produced, one for each year the task was run: CLEF2017, CLEF2018 and CLEF2019. The first two data sets contained exclusively Diagnostic Test Accuracy reviews (the goal of which is to determine the effectiveness of some medical diagnosis method). The CLEF2019 data set extended this to several other review types: Intervention, Prognosis and Qualitative. Following \citet{li2020stop}, only the Diagnostic Test Accuracy reviews are used for the experiments reported here, yielding 30 reviews\footnote{The CLEF2017 data set contained a total of 42 Diagnostic Test Accuracy reviews with 12 used to train algorithms.} 
(topics) from CLEF2017, 30 from CLEF2018 and 31 from CLEF2019.

Each topic in the CLEF2017, CLEF2018 and CLEF2019 data sets was derived from a systematic review produced by the Cochrane Collaboration.\footnote{\url{https://www.cochranelibrary.com/}} The document collection was the Medline database containing abstracts of scientific publications in the life sciences and associated fields. Topics consist of a topic/review title, a Boolean query developed by Cochrane experts and the set of PubMed Document Identifiers (PMIDs) returned by running the query over Medline. The goal of the task is to identify PMIDs of scientific papers that were included in the review, a time consuming task that is normally carried out manually. The topic titles are generally significantly longer and contain more technical terminology than those normally submitted to search engines. 

{\bf TREC Total Recall}\footnote{Available from \url{http://plg.uwaterloo.ca/~gvcormac/total-recall/}} The goal of the TREC Total Recall Track is to assess TAR methods with a human assessor forming part of the retrieval process (so the ground truth document relevance is revealed for each document immediately following its retrieval) and which aims to achieve very high recall (as close 100\% as possible). Following \citet{li2020stop}, the {\it athome4} dataset from the TREC 2016 Total Recall track \cite{grossman2016trec} is used to test approaches. This data set consists of 34 topics. 

The document collection for the data set consists of 290,099 emails from Jeb Bush’s eight-year tenure as Governor of Florida that was also used for the previous year's Total Recall exercise \cite{roegiest2015trec}. Each topic is based on an issue associated with Jeb Bush's governorship, e.g. {\it Felon disenfranchisement} and {\it Bottled Water}. Topics consist of a short title, normally a few words long and similar to the queries typically submitted to search engines (e.g. {\it Olympics}), and a slightly longer textual description (e.g. {\it Bid to host the Olympic games in Florida}). 

{\bf TREC Legal}\footnote{Available from \url{https://trec-legal.umiacs.umd.edu/}} The TREC Legal track \cite{cormack2010overview} focuses on TAR in the eDiscovery process where the aim is to identify (nearly) all documents relevant to a request for production in civil litigation while minimising the number of non-relevant documents examined. Topics {\it 303} and {\it 304} from the interactive task of the TREC 2010 Legal track are used. 

The document collection is a version of the ENRON data set based on the emails captured and made public by the Federal Energy Review Commission as part of their investigation into the collapse of Enron. This version contains 685,592 documents made up from 455,449 email messages and  230,143 attachments. Topics in this data set take the form of mock legal complaints that request disclosure of documents containing specific information (e.g. topic {\it 303} requests documents containing information related to the lobbying of public officials). In addition, topic {\it 304} is a ``privilege'' topic intended to model a search for documents that could be withheld from a production request on the basis of legal privilege.

\subsection{Rankings}\label{sec:rankings}

Stopping methods operate over a ranking of documents in a collection. Some approaches have chosen to closely integrate the stopping method with the ranking process (e.g. \cite{cormack2016scalability,li2020stop,hollmann2017ranking,Yang2021heuristic}) while others, including the one presented here, can be applied to any ranking of the collection (e.g. \cite{cormack2016engineering,Howard2020,callaghan2020statistical}). The goals of the evaluation include comparing the proposed approach against existing methods and determining how robust approaches are under a range of rankings. Ideally, it would have been possible to evaluate the approaches against multiple rankings for each data set, however, these are not always available, and evaluation was carried out using two sets of rankings: 

\begin{itemize}
    \item The first ranking used for the evaluation is produced by the AutoStop system \cite{li2020stop}. The AutoStop ranking algorithm is a CAL approach based on AutoTAR \cite{cormack2016scalability} and represents state-of-the-art performance. The reference implementation of AutoStop provided by \citet{li2020stop} was used to provide a ranking for each of the datasets used in the evaluation. These rankings allow the approaches to be evaluated and directly compared with existing approaches on multiple datasets. This ranking is used for the experiments reported in Sections \ref{sec:baseline}, \ref{sec:rate_comparison}, \ref{sec:topics_comparison}, \ref{sec:rel_docs}, \ref{sec:confidence_comparison} and \ref{sec:batches}.

    \item The second set of rankings were produced by participants in the CLEF task Technology-Assisted Review in Empirical Medicine \cite{kanoulas2017clef,kanoulas2018clef}. Rankings produced by systems that took part in the evaluations were made available by the task organisers.\footnote{Available from \url{https://github.com/CLEF-TAR}} The description of the CLEF2017 evaluation \cite{kanoulas2017clef} 
    states that 33 of the runs made available ranked the full set of documents returned by the Boolean query, however, four of these appear to contain fewer documents than the others and were therefore excluded from the experiments. Similarly, for the CLEF2018 task, 22 rankings were made available but documents were missing from eleven of these, leaving the remaining eleven for the experiments.\footnote{The CLEF2018 task placed more emphasis on identifying stopping points than its predecessor, which may explain the increased portion of submissions that did not attempt to rank all of the documents returned by the Boolean query.} (Rankings from CLEF2019 were not included in the experiments, given the small number of participants in the task's final iteration and the limited number of rankings available.) Results of the CLEF Empirical Medicine evaluations revealed that the rankings submitted varied considerably in their effectiveness, which is to be expected since the submissions ranged from applications of state-of-the-art approaches to experimental systems and (in the case of two runs) baseline approaches designed to provide context. These rankings can therefore be used to explore how the stopping approaches are affected by ranking effectiveness. Results of experiments using these rankings can be found in Section \ref{sec:multiple_rankings}. It is worth mentioning that previous stopping methods have been evaluated against single rankings. Evaluation using the multiple rankings available for this dataset provides valuable insight into the relationship between ranking and stopping effectiveness.

\end{itemize}

\section{Results}

\subsection{Baseline Comparison}\label{sec:baseline}

\subsubsection{Hyperparameter Tuning}\label{sec:hyperparameter}

The approach proposed in Section \ref{sec:algorithm} includes hyperparameters for which values have to be chosen before it can be compared against baseline methods. To ensure a fair comparison, values for these hyperparameters were selected by carrying out a grid search over the training portion of the CLEF 2017 dataset for three different levels of target recall: 1.0, 0.9 and 0.8. A single set of hyperparameters was used for all experiments. While it would have been possible to select a different set of hyperparameters for each dataset, potentially improving performance, doing so would have produced a less generalised model. 

The following hyperparameters were included in the grid search: counting process model $\in$ \{Inhomogeneous Poisson (IP), Cox\}, rate function $\in$ \{exponential, Power Law, Hyperbolic, AP-Prior\}, threshold for NRMSE fit $\in$ \{0.05, 0.1, 0.15\} and minimum number of relevant documents in the sample $\in$  \{10, 20, gradient decreasing\}. The $\alpha$ and $\beta$ parameters were not included in the grid search to reduce the computational cost required, and since altering them appeared to have limited effect on performance. Both were set to 0.025, values which lead the algorithm to check whether to stop at regular small intervals. Selecting the best hyperparameter values is not straightforward since the stopping problem has multiple objectives (i.e. achieving target recall while minimising the number of documents examined). For each target recall, the set of configurations that formed the Pareto frontier were identified. Hyperparameter values that appeared most frequently in this set were then chosen with the following selection as follows, counting process model: Inhomogeneous Poisson, rate function: hyperbolic, NRMSE threshold: 0.1, 
minimum number of relevant documents in the sample: gradient decreasing. The confidence parameter ($p$) was set to 0.95 for all experiments except those in Section \ref{sec:confidence_comparison}.

Overall, the choice of hyperparameter had a limited effect on performance. The most significant ones were the choice of rate function (explored later in Section \ref{sec:rate_comparison}) and the minimum number of relevant documents in the sample. For the second of these,  the flexibility of the gradient decreasing approach appeared to be useful to help the approach adapt to the varying number of relevant documents across topics.

\subsubsection{Comparison with Alternative Approaches}\label{sec:comparison}

\begin{figure}[h]
  \centering
\includegraphics[width=\textwidth,height=\textheight,keepaspectratio=true,trim=0 0 0 0, clip]{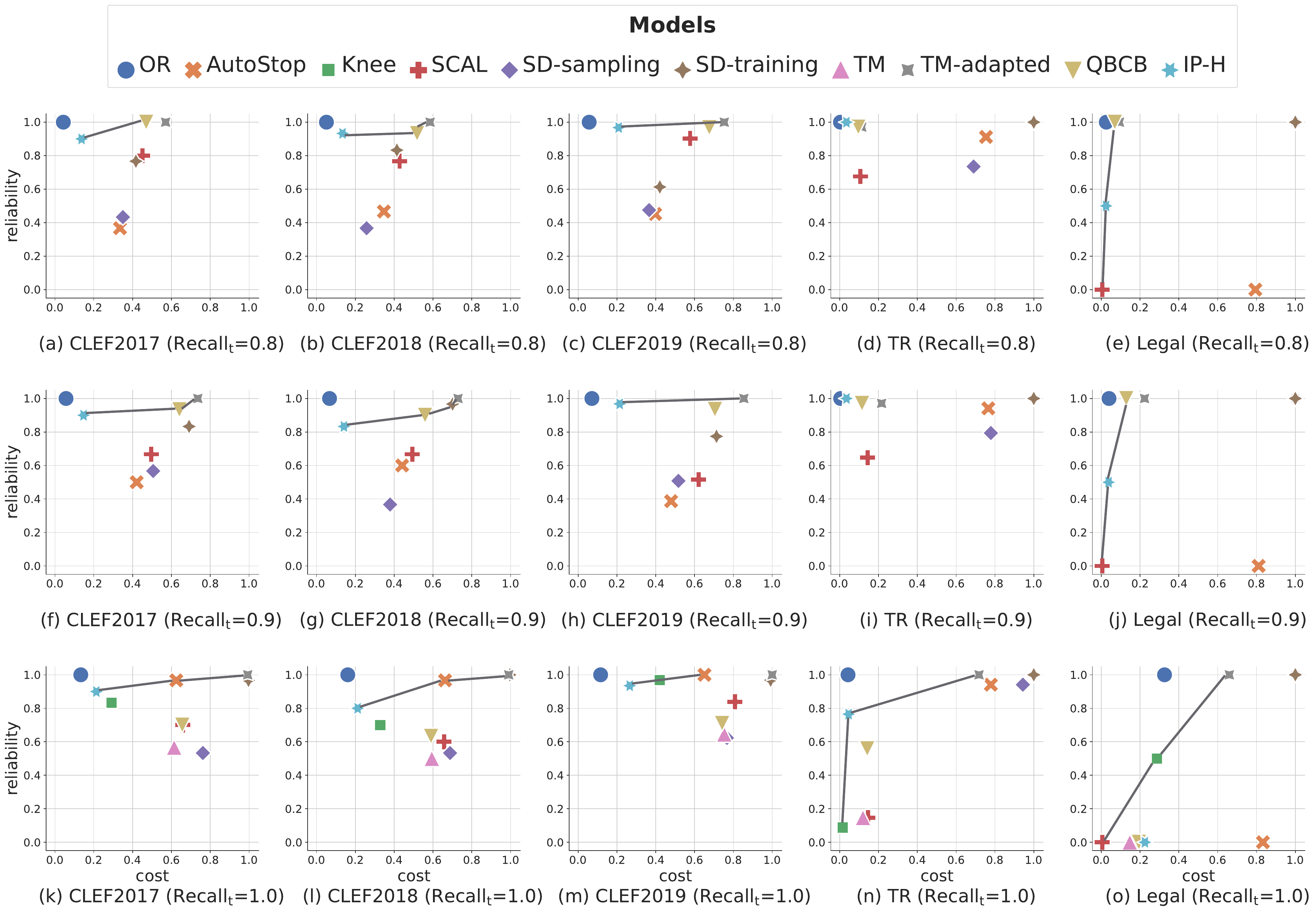}

  \caption{Cost vs. reliability for a range of approaches on multiple datasets. Pareto optimal points are linked by a grey line.}\label{fig:all_grid_col}
\end{figure}

Figure \ref{fig:all_grid_col} compares the performance of our approach against the various baselines described in Section \ref{sec:baselines}. Results for the majority of methods are those reported by \citet{li2020stop}. The exceptions are our own approach, the oracle, QBCB and adapted target methods.\footnote{It is not possible to set the target recall to 1.0 using the adjusted target method (see Section \ref{sec:baseline_target}). Instead, the target recall was set to 0.99 which restricted the number of relevant documents to a reasonable number (i.e. 300). Increasing the target recall further would have required a larger number of relevant documents to be found, e.g. a target recall of 0.999 would require 2996, and often more than the number of relevant documents in the collection.} More detailed results, including additional metrics, are provided in Tables \ref{tab:0.99recall_all}, \ref{tab:0.9recall_all} and \ref{tab:0.8recall_all} (Appendix \ref{sec:detailed_baseline_results}).

Figure \ref{fig:all_grid_col} shows the results for various target recalls (0.8, 0.9, 1.0) 
 along each row with different datasets (CLEF2017, CLEF2018, CLEF2019, TREC Total Recall and TREC Legal) in each column. Performance of the oracle approach (shown as a blue circle) indicates the minimum number of documents that need to be examined to reach the target recall. The oracle's reliability is always 1.0 since this approach is guaranteed to achieve the target recall. 

Comparing the results over all configurations (datasets and target recalls), the proposed model (denoted by a cyan star) performs well in terms of balancing reliability and cost. It was able to achieve the target recall with high reliability and lower cost than other approaches the majority of the time. The proposed approach is also Pareto optimal in the majority of cases, and is the only Pareto optimal approach in two cases: Total Recall dataset with target recall 0.8 or 0.9 (Figures \ref{fig:all_grid_col}d and \ref{fig:all_grid_col}i). 
The only case where our approach is not Pareto optimal occurs with the TREC Legal dataset when the target recall 1.0 (Figure \ref{fig:all_grid_col}o). The reliability scores for our approach are low for this dataset. However, reliability scores are also low for several other approaches and the overall pattern of results is somewhat different compared with other datasets. The proposed model was very close to reaching the target recall of 1.0 for one of the two topics in this collection (the recall achieved was 0.999) and would also have been Pareto optimal in this case if it had been able to identify the last few relevant documents. Analysis of the rankings for this dataset showed that the majority of the relevant documents were found very quickly, but one topic also contained a long tail of relevant documents (leading to the high oracle cost when the target recall is 1.0, see Figure \ref{fig:all_grid_col}o). 
The hyperbolic rate function produced a reasonable fit to the true rate during the early part of this ranking but underestimated the rate at which relevant documents occurred later, leading to premature stopping.
In fact, for this topic the proposed approach stopped at the earliest opportunity, after only the initial sample of 2.5\% of the documents had been analysed, which could potentially have been avoided by increasing the initial sample size, effectively applying a heuristic to say that stopping should only be considered after a certain portion of the documents have been examined. Choosing a different rate function increased reliability on this data set, albeit at increased cost (see Section \ref{sec:rate_comparison}).

\subsubsection{Target Set Methods}

The adapted target model and QBCB methods are also Pareto optimal in many cases. The approach used by these two methods is very similar since they both rely on a target set of relevant documents and stop when the last of these has been found in the ranking. These target set methods outperform the proposed approach in terms of reliability but not cost. In fact, the number of extra documents that have to be examined by this approach is often considerable and in some cases the entire collection. The most likely reason for this higher cost is that target methods do not take account of the fact that the likelihood of observing relevant documents decreases later in the ranking, leading them to sample high numbers of non-relevant documents. 

The difference between the performance of the two target set methods is most pronounced for target recall 1.0. The adapted target method is reliable but has a very high cost, requiring all documents to be examined for most collections. While the cost for the QBCB method is much lower, reliability substantially reduces (although results in Table \ref{tab:0.99recall_all} show that the recall is close to the target). 

Figure \ref{fig:all_grid_col} shows that the difference between the cost of these methods and the oracle varies between collections. It is highest for the smallest collections (the three CLEF collections) and lowest for the largest collection, TREC Legal. A possible explanation for this pattern is that larger collections provide the opportunity for more accurate estimates of the number of relevant documents during the random sampling used to create the target set.

Figure \ref{fig:all_grid_col} also allows comparison of both the original and adapted versions of the target method for target recall 1.0 (represented respectively as a pink triangle and grey cross). The adapted target method is more reliable than the approach used by \citet{li2020stop}. This is perhaps unsurprising since the statistical theory behind the approach requires an appropriate number of relevant documents to be found in order to provide theoretical guarantees about the recall levels achieved. On the other hand, the adapted version is more costly (due to the increased target size).

\subsection{Comparison of Rate Functions and Point Processes}\label{sec:rate_comparison}

One of the goals of this work is to compare the various rate functions and point processes described in Section \ref{sec:rate_function}. This was explored by running the proposed approach with each rate function using both the Inhomogeneous Poisson and Cox processes while fixing all other hyperparameters to the values described in Section \ref{sec:hyperparameter}.

\begin{table}[tp]
    \centering
    \caption{Comparison of performance of rate functions for 0.9 target recall. $\uparrow$ and $\downarrow$ indicate metrics where higher and lower scores are preferred (respectively). IP = Inhomogeneous Poisson Process, CX = Cox Process, P = power law, H = hyperbolic, E = exponential and A = AP Prior, e.g. ``CX-P'' indicates a Cox Process with the power law rate function.}
    \label{tab:IP_CX_all_rates}
    \resizebox*{!}{0.9\textheight}{%
    \begin{tabular}{c|c|cccccc}
    \toprule
    
        Dataset & Model & recall ($\uparrow$) & cost ($\downarrow$) & reliability ($\uparrow$)  & loss\textsubscript{er} ($\downarrow$) & RE ($\downarrow$) \\ 
        \midrule
        & IP-P & 1.000 ± 0.000 & 0.281 ± 0.226 & 1.000 & 0.030 ± 0.031 & 0.111 ± 0.000 \\ 
        & CX-P & 1.000 ± 0.000 & 0.278 ± 0.220 & 1.000  & 0.030 ± 0.031 & 0.111 ± 0.000 \\ 
        & IP-H & 0.955 ± 0.119 & 0.147 ± 0.114 & 0.900 & 0.036 ± 0.055 & 0.123 ± 0.076 \\ 
        \multirow{2}{*}{CLEF 2017} & CX-H & 0.951 ± 0.137 & 0.172 ± 0.140 & 0.900  & 0.044 ± 0.071 & 0.130 ± 0.096 \\
        & IP-E & 0.984 ± 0.031 & 0.154 ± 0.113 & 0.967 & 0.022 ± 0.032 & 0.094 ± 0.034 \\ 
        & CX-E & 0.984 ± 0.031 & 0.153 ± 0.112 & 0.967  & 0.022 ± 0.032 & 0.094 ± 0.034 \\ 
        & IP-A & 0.994 ± 0.017 & 0.205 ± 0.153 & 1.000 & 0.025 ± 0.032 & 0.105 ± 0.019 \\ 
        & CX-A & 0.994 ± 0.017 & 0.205 ± 0.153 & 1.000  & 0.025 ± 0.032 & 0.105 ± 0.019 \\ 
        \midrule
        
        & IP-P & 1.000 ± 0.001 & 0.293 ± 0.213 & 1.000 & 0.024 ± 0.026 & 0.111 ± 0.001 \\
        & CX-P & 1.000 ± 0.001 & 0.292 ± 0.211 & 1.000  & 0.024 ± 0.026 & 0.111 ± 0.001 \\
        & IP-H & 0.941 ± 0.125 & 0.141 ± 0.145 & 0.833 & 0.030 ± 0.063 & 0.115 ± 0.088 \\
        \multirow{2}{*}{CLEF 2018} & CX-H & 0.948 ± 0.126 & 0.167 ± 0.148 & 0.867  & 0.031 ± 0.062 & 0.120 ± 0.087 \\
        & IP-E & 0.978 ± 0.034 & 0.169 ± 0.175 & 0.933 & 0.015 ± 0.020 & 0.091 ± 0.027 \\ 
        & CX-E & 0.977 ± 0.034 & 0.167 ± 0.176 & 0.933  & 0.015 ± 0.020 & 0.089 ± 0.028 \\ 
        & IP-A & 0.988 ± 0.029 & 0.221 ± 0.189 & 0.967 & 0.018 ± 0.021 & 0.099 ± 0.028 \\ 
        & CX-A & 0.988 ± 0.029 & 0.221 ± 0.189 & 0.967  & 0.018 ± 0.021 & 0.099 ± 0.028 \\ 
        \midrule
        
        & IP-P & 0.999 ± 0.006 & 0.283 ± 0.209 & 1.000 & 0.047 ± 0.061 & 0.110 ± 0.007 \\
        & CX-P & 0.999 ± 0.006 & 0.282 ± 0.207 & 1.000  & 0.047 ± 0.061 & 0.110 ± 0.007 \\ 
        & IP-H & 0.984 ± 0.061 & 0.214 ± 0.177 & 0.968 & 0.043 ± 0.063 & 0.110 ± 0.032 \\ 
        \multirow{2}{*}{CLEF 2019} & CX-H & 0.984 ± 0.062 & 0.232 ± 0.177 & 0.968  & 0.046 ± 0.063 & 0.111 ± 0.034 \\ 
        & IP-E & 0.989 ± 0.032 & 0.215 ± 0.176 & 0.968 & 0.040 ± 0.062 & 0.104 ± 0.014 \\
        & CX-E & 0.989 ± 0.032 & 0.214 ± 0.176 & 0.968  & 0.040 ± 0.062 & 0.104 ± 0.015 \\ 
        & IP-A & 0.993 ± 0.031 & 0.247 ± 0.180 & 0.968 & 0.044 ± 0.061 & 0.109 ± 0.009 \\
        & CX-A & 0.993 ± 0.031 & 0.247 ± 0.180 & 0.968  & 0.044 ± 0.061 & 0.109 ± 0.009 \\
        \midrule
        
        & IP-P & 1.000 ± 0.001 & 0.059 ± 0.085 & 1.000 & 0.000 ± 0.000 & 0.111 ± 0.001 \\
        & CX-P & 1.000 ± 0.001 & 0.059 ± 0.085 & 1.000  & 0.000 ± 0.000 & 0.111 ± 0.001 \\ 
        & IP-H & 0.999 ± 0.002 & 0.035 ± 0.024 & 1.000 & 0.000 ± 0.000 & 0.110 ± 0.002 \\
        \multirow{2}{*}{TR} & CX-H & 1.000 ± 0.001 & 0.043 ± 0.034 & 1.000  & 0.000 ± 0.000 & 0.111 ± 0.001 \\ 
        & IP-E & 0.999 ± 0.002 & 0.030 ± 0.022 & 1.000 & 0.000 ± 0.000 & 0.110 ± 0.002 \\
        & CX-E & 0.999 ± 0.002 & 0.030 ± 0.022 & 1.000  & 0.000 ± 0.000 & 0.110 ± 0.002 \\
        & IP-A & 1.000 ± 0.001 & 0.041 ± 0.048 & 1.000 & 0.000 ± 0.000 & 0.111 ± 0.001 \\ 
        & CX-A & 1.000 ± 0.001 & 0.041 ± 0.048 & 1.000  & 0.000 ± 0.000 & 0.111 ± 0.001 \\ 
        \midrule
        
        & IP-P & 1.000 ± 0.001 & 0.425 ± 0.035 & 1.000 & 0.002 ± 0.001 & 0.111 ± 0.001 \\
        & CX-P & 1.000 ± 0.001 & 0.412 ± 0.053 & 1.000  & 0.002 ± 0.001 & 0.111 ± 0.001 \\
        & IP-H & 0.846 ± 0.148 & 0.038 ± 0.018 & 0.500 & 0.035 ± 0.046 & 0.117 ± 0.085 \\
        \multirow{2}{*}{Legal} & CX-H & 0.793 ± 0.074 & 0.025 ± 0.000 & 0.000  & 0.045 ± 0.030 & 0.119 ± 0.082 \\
        & IP-E & 0.930 ± 0.030 & 0.050 ± 0.000 & 1.000 & 0.005 ± 0.004 & 0.033 ± 0.033 \\
        & CX-E & 0.877 ± 0.045 & 0.038 ± 0.018 & 0.500  & 0.016 ± 0.011 & 0.035 ± 0.036 \\
        & IP-A & 0.793 ± 0.074 & 0.025 ± 0.000 & 0.000 & 0.045 ± 0.030 & 0.119 ± 0.082 \\
        & CX-A & 0.793 ± 0.074 & 0.025 ± 0.000 & 0.000  & 0.045 ± 0.030 & 0.119 ± 0.082 \\ 

        \bottomrule
    \end{tabular}
   }
\end{table}

\begin{table}[]
    \centering
    \caption{Statistical significance of differences between performance of four rate functions (P = power law, H = hyperbolic, E = exponential and A = AP Prior) for each metric (paired t-test with Bonferroni correction, p < 0.05). \ding{51} indicates a difference is significant and \ding{55} otherwise. Significance is computed using data from all datasets and recall levels used in experiments.}
    \label{tab:stat_sig}
    \begin{tabular}{c|ccccc}
    \hline
      Rate comparison    & recall & cost & reliability & loss\textsubscript{er} & RE\\
      \hline
       H vs. P &  \ding{51} & \ding{51} & \ding{51} & \ding{51} & \ding{55} \\
        H vs. E & \ding{51} & \ding{55} & \ding{55} & \ding{51} & \ding{51} \\
        H vs. A & \ding{55} & \ding{55} & \ding{55} & \ding{51} & \ding{55} \\
        P vs. E & \ding{51} & \ding{51} & \ding{51} & \ding{55} & \ding{55}\\
        P vs. A & \ding{51} & \ding{51} & \ding{51} & \ding{55} & \ding{55}\\
        E vs. A & \ding{55} & \ding{55} & \ding{55} & \ding{55} & \ding{55}\\
    \hline
    \end{tabular}
\end{table}

Table \ref{tab:IP_CX_all_rates} shows results for a target recall of 0.9. (Similar patterns of results were observed for different target recalls.) Scores for the five metrics described in Section \ref{sec:metrics} are shown followed by standard deviation across topics in the relevant collection. (Note that standard deviation is not included for the reliability metric since, unlike the other metrics, it is defined across all topics in a collection rather than each topic individually.) Results show how the behaviour of the proposed approach varies according to the rate function applied. 
Statistical significance of the difference in performance of the four rate functions is shown in Table \ref{tab:stat_sig}.

Overall, the hyperbolic decline rate achieved the target recall with minimal cost and highest reliability in the majority of cases. The power law rate function was the most reliable but also has the highest cost. Performance of the other two rate functions lies between that of the hyperbolic and power law.

For the majority of the collections all variants achieve average recall above (or very close to) the target and achieve the target recall with high reliability. For the three CLEF collections, this is always achieved by examining no more than one third of the collection and in many cases substantially less. For the Total Recall collection, the recall and reliability are high (with near-perfect recall) with a very low cost. All relevant documents are identified while only examining (at most) 6\% of the documents. However, the Legal collection presents more of a challenge to the approach and the effect of varying the rate function is more pronounced. The recall and reliability are higher for the power law but only at the cost of requiring an order of magnitude more documents to be examined. This difference in the performance of the various rate functions is likely due to the rankings produced for this collection (see Section \ref{sec:comparison}) since the rate functions fitted to these have the potential to decrease very rapidly but the actual rate at which this happens depends upon the particular function being used.

Table \ref{tab:IP_CX_all_rates} also highlights the similarity between the results produced using the Inhomogeneous Poisson and Cox processes. 
The differences between the results produced by the two processes were found to be statistically significant for all metrics with the exception of cost (paired t-test, p < 0.05). 
On average, the Inhomogeneous Poisson process achieved higher recall and reliability and lower relative error than the Cox process. Although the Inhomogeneous Poisson process had a higher cost than the Cox process, the difference between them was not statistically significant. The Cox Process is more computationally expensive than the Inhomogeneous Poisson Process since the integral over the potential parameters of the rate function (see Section \ref{sec:cox}) cannot be expressed as a convenient closed form like the Poisson Process and, instead, is estimated using numerical integration. Given this trade-off, the Inhomogeneous Poisson Process may be preferable to the Cox Process in most circumstances and is used for the remainder of the experiments.

\subsection{Performance Across Topics}\label{sec:topics_comparison}

\begin{figure}[!htbp]
     \centering
     \begin{subfigure}[b]{0.30\textwidth}
         \centering
         
         \includegraphics[width=\textwidth]{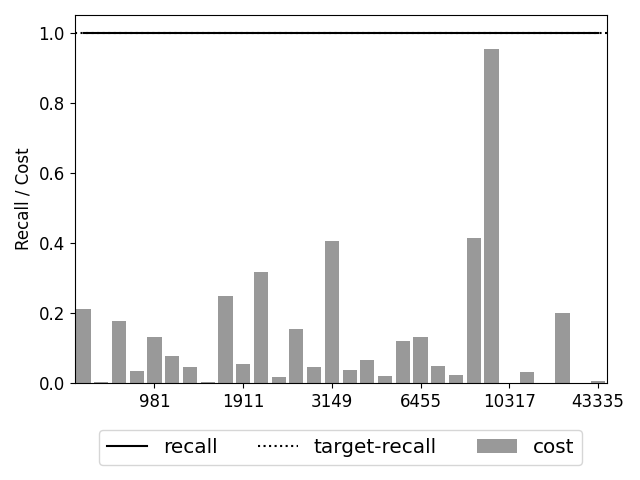}
         \caption{Oracle (Recall$_t$ = 1.0)}
         \label{fig:oracle_0.999}
     \end{subfigure}
     \begin{subfigure}[b]{0.30\textwidth}
         \centering
         \includegraphics[width=\textwidth]{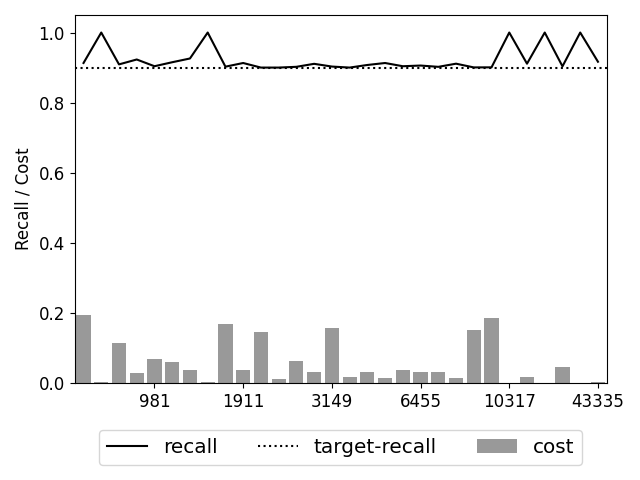}
         \caption{Oracle (Recall$_t$ = 0.9)}
         \label{fig:oracle_0.9}
     \end{subfigure}
     \begin{subfigure}[b]{0.30\textwidth}
         \centering
         \includegraphics[width=\textwidth]{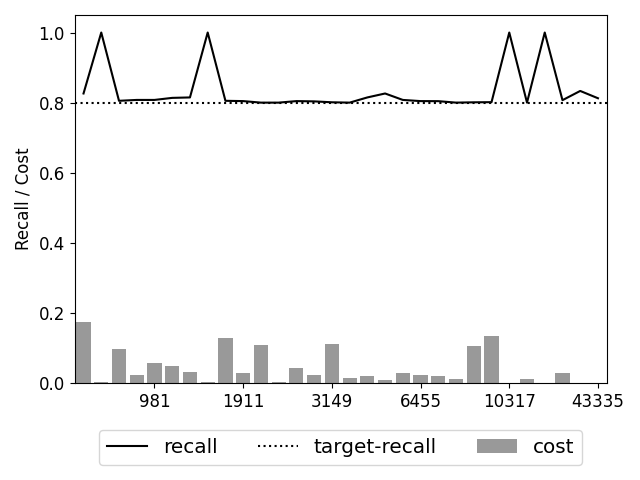}
         \caption{Oracle (Recall$_t$ = 0.8)}
         \label{fig:oracle_0.8}
     \end{subfigure}
     \begin{subfigure}[b]{0.30\textwidth}
         \centering
         \includegraphics[width=\textwidth,trim=0 40 0 0, clip]{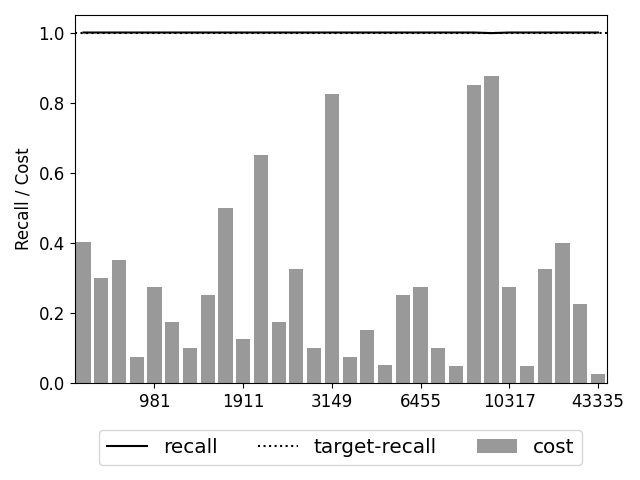}
         \caption{IP-P (Recall$_t$ = 1.0)}
         \label{fig:ip_p_0.999}
     \end{subfigure}
     \begin{subfigure}[b]{0.30\textwidth}
         \centering
         \includegraphics[width=\textwidth,trim=0 40 0 0, clip]{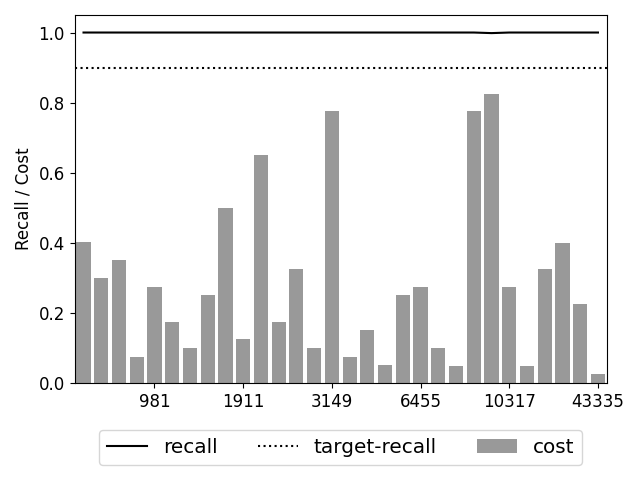}
         \caption{IP-P (Recall$_t$ = 0.9)}
         \label{fig:ip_p_0.9}
     \end{subfigure}
     \begin{subfigure}[b]{0.30\textwidth}
         \centering
         \includegraphics[width=\textwidth,trim=0 40 0 0, clip]{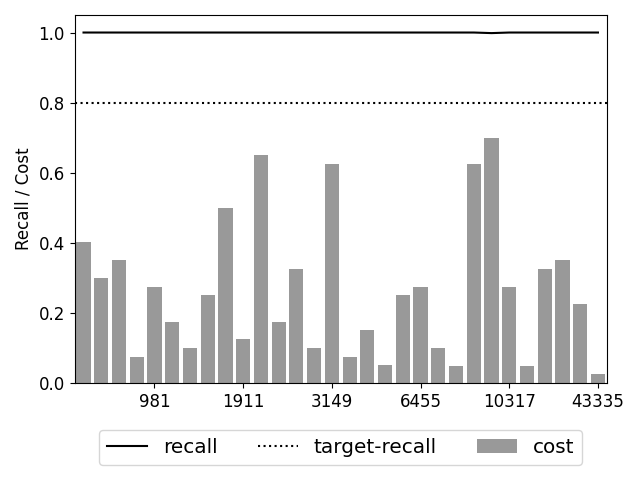}
         \caption{IP-P (Recall$_t$ = 0.8)}
         \label{fig:ip_p_0.8}
     \end{subfigure}
     \begin{subfigure}[b]{0.30\textwidth}
         \centering
         \includegraphics[width=\textwidth,trim=0 40 0 0, clip]{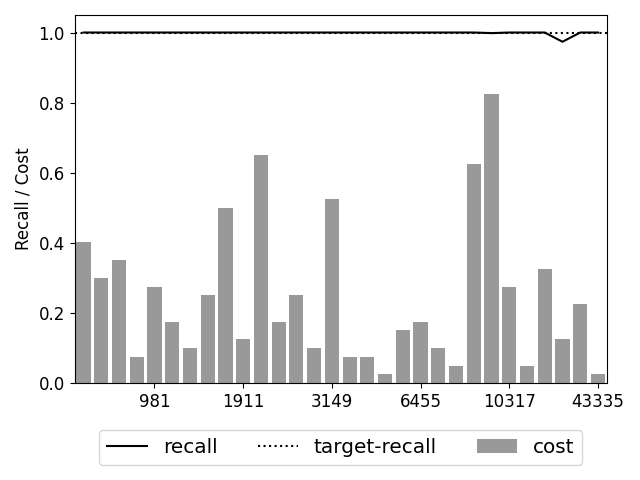}
         \caption{IP-E (Recall$_t$ = 1.0)}
         \label{fig:ip_e_0.999}
     \end{subfigure}
     \begin{subfigure}[b]{0.30\textwidth}
         \centering
         \includegraphics[width=\textwidth,trim=0 40 0 0, clip]{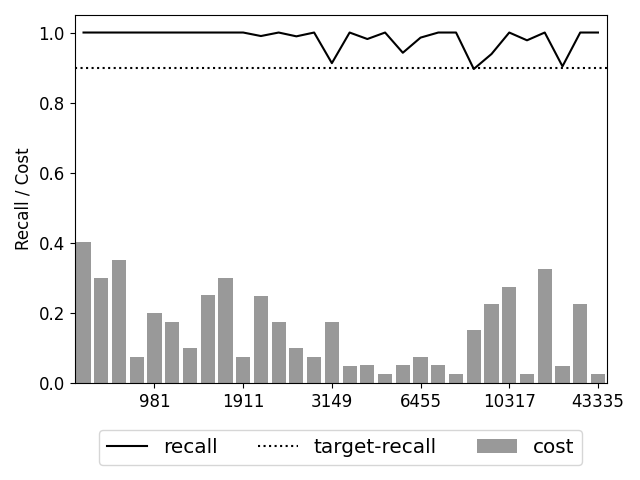}
         \caption{IP-E (Recall$_t$ = 0.9)}
         \label{fig:ip_e_0.9}
     \end{subfigure}
     \begin{subfigure}[b]{0.30\textwidth}
         \centering
         \includegraphics[width=\textwidth,trim=0 40 0 0, clip]{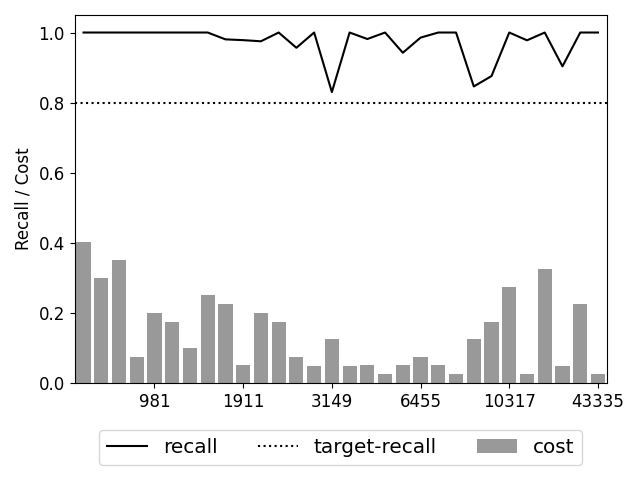}
         \caption{IP-E (Recall$_t$ = 0.8)}
         \label{fig:ip_e_0.8}
     \end{subfigure}
     \begin{subfigure}[b]{0.30\textwidth}
         \centering
         \includegraphics[width=\textwidth,trim=0 40 0 0, clip]{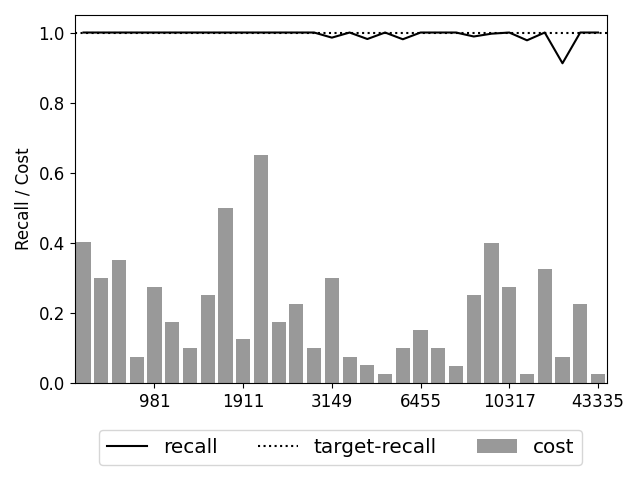}
         \caption{IP-A (Recall$_t$ = 1.0)}
         \label{fig:ip_a_0.999}
     \end{subfigure}
     \begin{subfigure}[b]{0.30\textwidth}
         \centering
         \includegraphics[width=\textwidth,trim=0 40 0 0, clip]{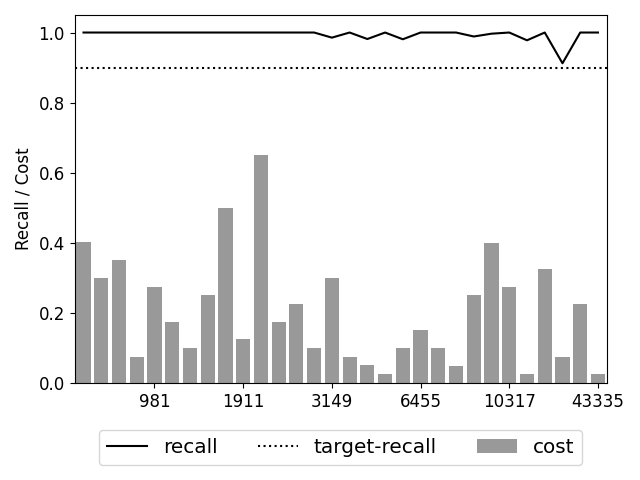}
         \caption{IP-A (Recall$_t$ = 0.9)}
         \label{fig:ip_a_0.9}
     \end{subfigure}
     \begin{subfigure}[b]{0.30\textwidth}
         \centering
         \includegraphics[width=\textwidth,trim=0 40 0 0, clip]{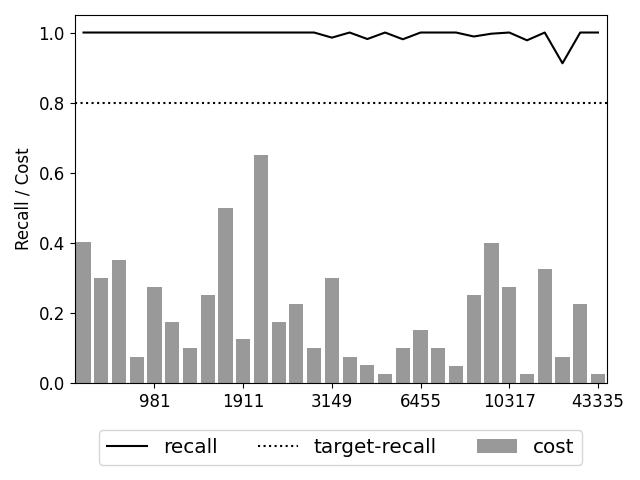}
         \caption{IP-A (Recall$_t$ = 0.8)}
         \label{fig:ip_a_0.8}
     \end{subfigure}
    \begin{subfigure}[b]{0.30\textwidth}
         \centering
         \includegraphics[width=\textwidth,trim=0 40 0 0, clip]{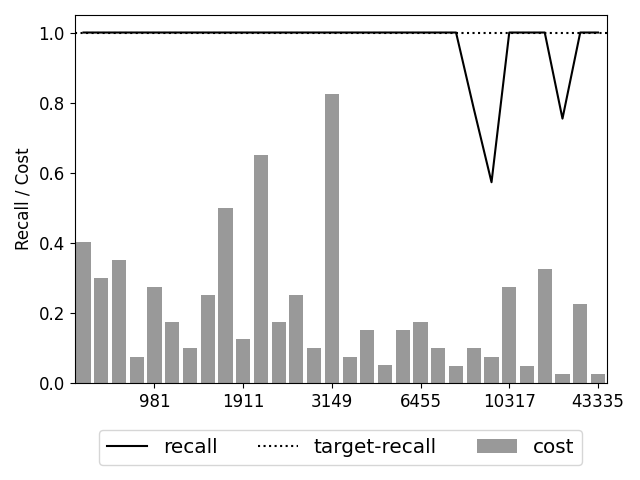}
         \caption{IP-H (Recall$_t$ = 1.0)}
         \label{fig:ip_h_0.999}
     \end{subfigure}
    \begin{subfigure}[b]{0.30\textwidth}
         \centering
         \includegraphics[width=\textwidth,trim=0 40 0 0, clip]{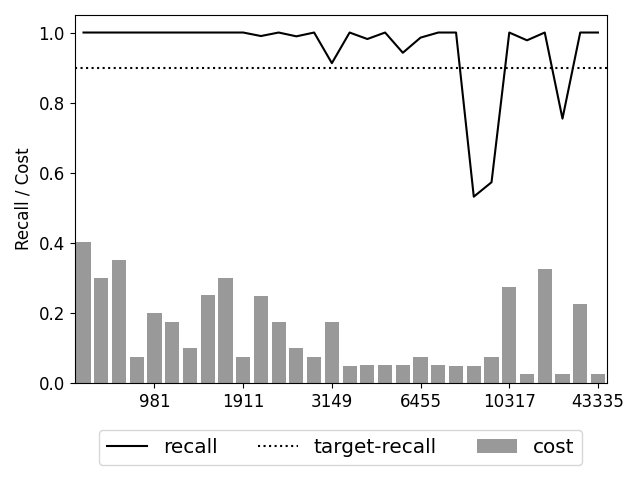}
         \caption{IP-H (Recall$_t$ = 0.9)}
         \label{fig:ip_h_0.9}
     \end{subfigure}
    \begin{subfigure}[b]{0.30\textwidth}
         \centering
         \includegraphics[width=\textwidth,trim=0 40 0 0, clip]{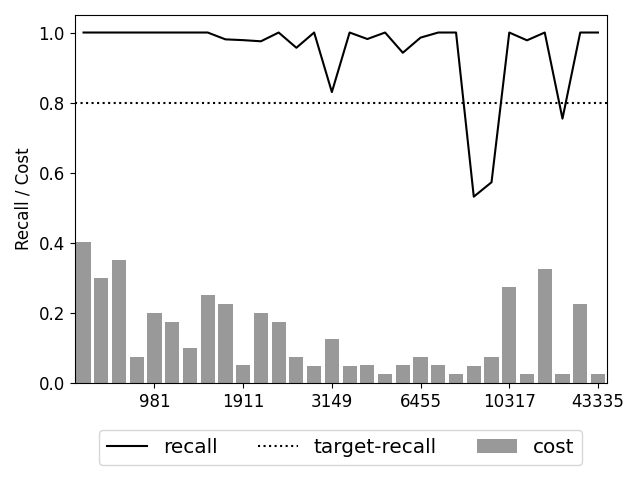}
         \caption{IP-H (Recall$_t$ = 0.8)}
         \label{fig:ip_h_0.8}
     \end{subfigure}

        \caption{Details of performance for each topic for CLEF 2017 collection. For each topic grey bars indicate the cost and black line represents the recall. The dotted horizontal line indicates the target recall. Topics are sorted by the number of documents they contain (ascending from left to right).}
        \label{fig:res_per_topics_clef2017}
\end{figure}

An analysis of performance across individual topics was also carried out with the results for the CLEF 2017 collection shown in Figure \ref{fig:res_per_topics_clef2017}. Results of the oracle method is shown in the top row (sub figures \ref{fig:oracle_0.999}, \ref{fig:oracle_0.9} and \ref{fig:oracle_0.8}) to provide context for the performance of the other approaches. The number of documents examined must be at least as high as the oracle cost to achieve the target recall. Results show that this number (indicated by a grey bar) varies considerably between topics when the target recall is set to 1 and almost all documents need to be examined for one topic (see Figure \ref{fig:oracle_0.999}). There is also a substantial drop in this number for lower target recalls and for many topics examining fewer than 20\% of the documents is sufficient to achieve a recall of 0.9 or  0.8 (see Figures \ref{fig:oracle_0.9} and \ref{fig:oracle_0.8}). 

Each column in Figure \ref{fig:res_per_topics_clef2017} shows performance obtained using the four rate functions for target recalls 1.0, 0.9 and 0.8. These figures reflect the overall pattern of results for the CLEF 2017 dataset shown in Table \ref{tab:IP_CX_all_rates}. For example, the power law is reliable but also has higher cost than other rate functions. Figures \ref{fig:ip_p_0.999}, \ref{fig:ip_p_0.9} and \ref{fig:ip_p_0.8} also show the algorithm is somewhat over cautious when this rate function is used since the cost is noticeably higher than for the oracle and other rate functions. In addition, there is little reduction in the number of documents examined when target recall is reduced. The other rate functions also tend to overshoot the target recall, although to a lesser extent than when the power law is used. The hyperbolic rate function is the only one which fails to reach the target recall for some topics (i.e. reliability < 1) for this dataset. Topics where the target recall is not achieved tend to be larger ones (shown towards the right of each figure) with the amount to which the recall falls short of the target recall varying by topic (see Figures \ref{fig:ip_h_0.999}, \ref{fig:ip_h_0.9} and \ref{fig:ip_h_0.8}). The set of topics for which the achieved recall falls short of the target is similar across the rate functions, suggesting that some topics are more problematic for the proposed approach than others. Additional analysis was carried out on the three topics where the target recall was not reached when the hyperbolic rate function was used (CD011975, CD011984 and CD010339). In all three rankings, the last relevant documents in the ranking were preceded by long sequences of irrelevant documents causing the rate function to underestimate the probability of finding relevant documents later in the ranking which, in turn, caused the algorithm to stop before the target recall was reached. 

Six of the topics in this dataset (around 20\% of the total) have a small number of relevant documents (between 1 and 10) which causes the algorithm to overshoot for all rate functions since it requires a minimum number of relevant documents to be identified before the considering stopping (see Section \ref{sec:pseudo}).

\subsection{Estimation of Number of Relevant Documents}\label{sec:rel_docs}

The next experiment assesses the accuracy of the estimation of the number of relevant documents remaining. Although determining this value is not the main goal of our approach, observing it provides useful information about its behaviour. 

\begin{figure}[!ht]
\centering
\includegraphics[width=0.8\textwidth]{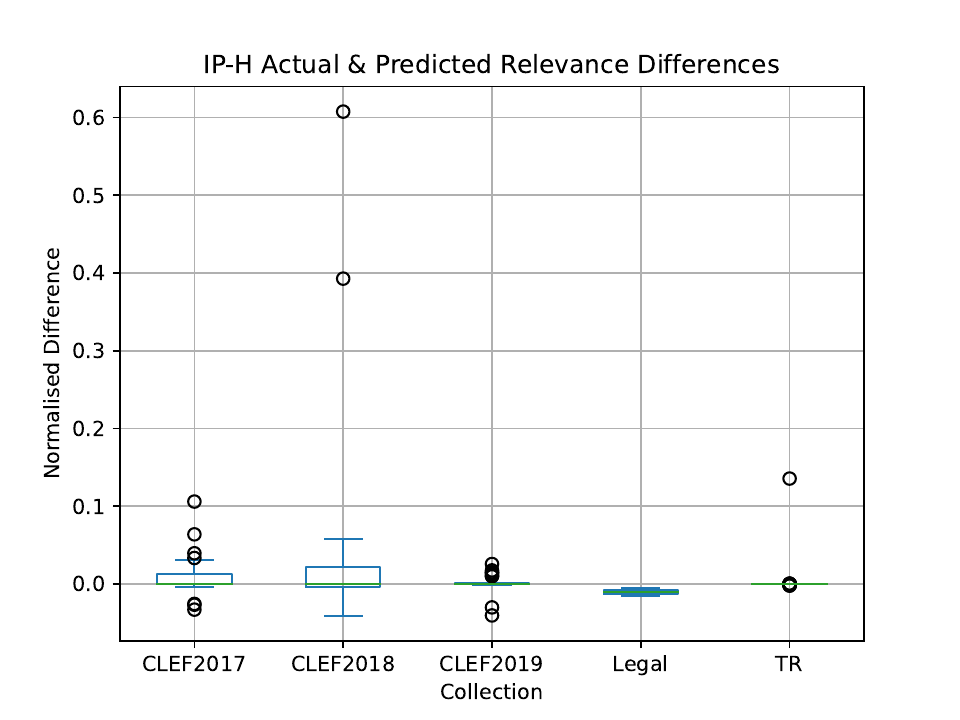}
\caption{Distribution of normalised differences between actual and predicted number of relevant documents remaining using IP-H. Boxes extend between the first and third quartiles with the median indicated by a green line. The whiskers expand the box by 1.5 $\times$ (Q3 - Q1), i.e. 1.5 times the inter-quartile range. Outliers beyond this range are indicated by circles.}

\label{fig:IP-H_Actual_Predicted}
\end{figure}

The normalised difference between the actual and predicted number of relevant documents remaining is calculated according to the following formula: 

\begin{equation}
    \frac{predicted - actual}{actual}
\end{equation}

where $predicted$ is the number of relevant documents remaining predicted by our approach and $actual$ is the actual number. The average of these values over all iterations using the IP-H approach is computed for each topic. Results are shown in Figure \ref{fig:IP-H_Actual_Predicted} where each collection is represented as a boxplot. 

The figure shows that the normalised difference is relatively small for the majority of topics, indicating that the estimation of the number of relevant documents remaining is broadly accurate in the majority of cases. Where the estimates are not accurate the model tends to overestimate which provides some explanation for why it sometimes overshoots the optimal stopping point. However, this is preferable to undershooting since our aim is to develop a method where the target recall is achieved. 

In some cases the overestimation is substantial, most notably in two topics in the CLEF2018 collection (CD011431 and CD008122). These topics were found to have a high prevalence of relevant documents and unusual patterns in the ranking where unusually high numbers of relevant documents appeared later in the ranking, meaning that the rate function fitted to the earlier part of the ranking did not provide a good indication of later behaviour.  

\begin{table}[t]
\centering
\caption{IP-H Model Stopping Performance Against Multiple Confidence Levels for CLEF 2017 dataset}\label{tab:confidence_clef2017}  
\begin{tabular}{c|ccccc}
\toprule
       $p$ & recall & cost & reliability  & loss\textsubscript{er} & RE \\ 
        \midrule					
 \multicolumn{6}{c}{Target Recall 1.0}  \\					
\midrule
 0.8 & 0.970 ± 0.095 & 0.202 ± 0.169 & 0.867  & 0.035 ± 0.042 & 0.030 ± 0.095 \\ 
0.6 & 0.969 ± 0.095 & 0.185 ± 0.143 & 0.800  & 0.033 ± 0.043 & 0.031 ± 0.095 \\ 
0.4 & 0.969 ± 0.095 & 0.180 ± 0.140 & 0.767  & 0.033 ± 0.043 & 0.031 ± 0.095 \\ 
0.2 & 0.968 ± 0.095 & 0.171 ± 0.135 & 0.767  & 0.032 ± 0.043 & 0.032 ± 0.095 \\ 
        \midrule					
 \multicolumn{6}{c}{Target Recall 0.9}  \\					
\midrule
0.8 & 0.955 ± 0.119 & 0.144 ± 0.114 & 0.900  & 0.036 ± 0.056 & 0.123 ± 0.076 \\ 
0.6 & 0.954 ± 0.119 & 0.143 ± 0.116 & 0.900  & 0.036 ± 0.056 & 0.122 ± 0.076 \\ 
0.4 & 0.953 ± 0.119 & 0.140 ± 0.114 & 0.867  & 0.036 ± 0.056 & 0.122 ± 0.076 \\ 
0.2 & 0.950 ± 0.121 & 0.138 ± 0.114 & 0.867  & 0.037 ± 0.055 & 0.123 ± 0.074 \\ 
        \midrule					
 \multicolumn{6}{c}{Target Recall 0.8}  \\					
\midrule
0.8 & 0.949 ± 0.120 & 0.137 ± 0.113 & 0.900  & 0.037 ± 0.055 & 0.231 ± 0.056 \\ 
0.6 & 0.947 ± 0.123 & 0.135 ± 0.112 & 0.867  & 0.037 ± 0.056 & 0.231 ± 0.057 \\ 
0.4 & 0.947 ± 0.123 & 0.135 ± 0.112 & 0.867  & 0.037 ± 0.056 & 0.231 ± 0.057 \\ 
0.2 & 0.947 ± 0.123 & 0.135 ± 0.112 & 0.867  & 0.037 ± 0.056 & 0.231 ± 0.057 \\ 
\bottomrule
\end{tabular}
\end{table}

\subsection{Effect of Varying Confidence Levels}\label{sec:confidence_comparison}

The proposed approach allows the confidence level ($p$) in the estimated total number of relevant documents to be varied. Experiments were carried out using a range of values for $p$ and target recall: $p \in \{0.8, 0.6, 0.4, 0.2 \}$ and target recall $\in \{ 1.0, 0.9, 0.8 \}$. Results for the CLEF 2017 data set are shown in Table \ref{tab:confidence_clef2017}. 
The effect of varying the value of $p$ is highest when the target recall is high and lowers as it is reduced. For a target recall of 1.0, reducing $p$ leads to a reduction in the cost and reliability with only a small reduction in recall, indicating that the approach is less conservative in deciding when to stop examining documents. However, the effect of varying $p$ is minor for a target recall of 1.0 and even smaller for lower target recalls. The reason for this limited effect is likely to be the steps taken to ensure that the algorithm does not stop too early because it has made predictions based on limited or unreliable evidence, e.g. few relevant documents or a badly fitted rate function (see Section \ref{sec:rate}), and may also be linked to the tendency to overestimate the number of relevant documents for some topics (see Section \ref{sec:rel_docs}). 

These results show that the reliability of our approach tends to exceed the confidence and remains high even when the confidence level is reduced. They also show that the $p$ parameter in our approach should not be interpreted in the same way as the confidence guarantees in some previous stopping algorithms, e.g. \cite{cormack2016engineering,li2020stop,callaghan2020statistical,lewis2021certifying}, where it can be interpreted as the proportion of cases in which the target recall will be reached (i.e. its reliability). The link between this probability and the algorithm's behaviour is less direct in our approach, although it still provides a mechanism through which it can be influenced.

\subsection{Performance on Multiple Rankings}\label{sec:multiple_rankings}

The next set of experiments explore the effect of ranking effectiveness on performance. The proposed approach is applied to the set of rankings made available for the CLEF 2017 and CLEF 2018 data sets (see Section \ref{sec:rankings}). Results were generated for target recalls 0.8 and 0.9 using each of the four rate functions, the Inhomogeneous Poisson Process and all other parameters set as they had been in previous experiments (see Section \ref{sec:hyperparameter}). 

Table \ref{tab:avrg_multiple_runs_clef2017_clef2018} shows the averaged (mean) performance and standard deviation over all runs in the relevant dataset. Results generated using the Oracle method are also included for comparison. Results show that, similar to the experiments carried out over a single ranking, the power law is the most reliable rate function but with the highest cost. Performance of the other three rate functions is broadly similar, although the average reliability across all runs is low.

\begin{table}[!t]
 \centering
    \caption{Averaged performance over multiple runs over CLEF 2017 and CLEF 2018 collections. 
    All differences between IP-$\ast$ and OR results are statistically significant (paired t-test, $p < 0.05$) with the exception of those indicated by an asterisk ($^*$).}
    \label{tab:avrg_multiple_runs_clef2017_clef2018}

 \begin{tabular}{l|l|llllll}
\toprule					
Dataset & Model & recall ($\uparrow$) & cost ($\downarrow$) & reliability ($\uparrow$) & loss\textsubscript{er} ($\downarrow$) & RE ($\downarrow$) \\ 
\midrule 
 &  & \multicolumn{5}{c}{Target Recall 0.9}   \\ 
 \midrule 
 & OR & 0.924 ± 0.000 & 0.471 ± 0.209 & 1.000 ± 0.000 & 0.173 ± 0.115 & 0.027 ± 0.000 \\ 
 & IP-P & 0.923 ± 0.054$^*$ & 0.692 ± 0.117 & 0.840 ± 0.080 & 0.299 ± 0.115 & 0.137 ± 0.040 \\ 
CLEF 2017 & IP-H & 0.699 ± 0.171 & 0.307 ± 0.050 & 0.354 ± 0.200 & 0.261 ± 0.161 & 0.284 ± 0.162 \\ 
 & IP-E & 0.832 ± 0.066 & 0.444 ± 0.168$^*$ & 0.549 ± 0.123 & 0.231 ± 0.131 & 0.164 ± 0.065 \\ 
 & IP-A & 0.759 ± 0.144 & 0.382 ± 0.069 & 0.457 ± 0.179 & 0.252 ± 0.151 & 0.240 ± 0.140 \\ 
\midrule 
 & OR & 0.912 ± 0.000 & 0.527 ± 0.261 & 1.000 ± 0.000 & 0.154 ± 0.120 & 0.013 ± 0.000 \\ 
 & IP-P & 0.911 ± 0.065$^*$ & 0.684 ± 0.083 & 0.755 ± 0.165 & 0.195 ± 0.063$^*$ & 0.127 ± 0.025 \\ 
CLEF2018 & IP-H & 0.663 ± 0.136 & 0.231 ± 0.031 & 0.245 ± 0.160 & 0.220 ± 0.096 & 0.299 ± 0.127 \\ 
 & IP-E & 0.769 ± 0.104 & 0.334 ± 0.087 & 0.357 ± 0.237 & 0.156 ± 0.058$^*$ & 0.197 ± 0.083 \\ 
 & IP-A & 0.749 ± 0.116 & 0.339 ± 0.057 & 0.364 ± 0.188 & 0.177 ± 0.076$^*$ & 0.232 ± 0.099 \\ 
\midrule 
 &  & \multicolumn{5}{c}{Target Recall 0.8}  \\ 
\midrule 
 & OR & 0.830 ± 0.000 & 0.356 ± 0.205 & 1.000 ± 0.000 & 0.140 ± 0.100 & 0.038 ± 0.000 \\ 
 & IP-P & 0.908 ± 0.067 & 0.633 ± 0.108 & 0.872 ± 0.086 & 0.265 ± 0.108 & 0.227 ± 0.024 \\ 
CLEF 2017 & IP-H & 0.683 ± 0.167 & 0.286 ± 0.052 & 0.446 ± 0.258 & 0.261 ± 0.162 & 0.298 ± 0.136 \\ 
 & IP-E & 0.785 ± 0.076 & 0.382 ± 0.157$^*$ & 0.604 ± 0.177 & 0.223 ± 0.122 & 0.215 ± 0.055 \\ 
 & IP-A & 0.759 ± 0.144 & 0.382 ± 0.069$^*$ & 0.601 ± 0.236 & 0.252 ± 0.151 & 0.277 ± 0.119 \\ 
\midrule 
 & OR & 0.811 ± 0.000 & 0.397 ± 0.234 & 1.000 ± 0.000 & 0.134 ± 0.095 & 0.014 ± 0.000 \\ 
 & IP-P & 0.886 ± 0.088 & 0.614 ± 0.076 & 0.800 ± 0.195 & 0.173 ± 0.060$^*$ & 0.202 ± 0.042 \\ 
CLEF 2018 & IP-H & 0.637 ± 0.132 & 0.200 ± 0.029 & 0.333 ± 0.243 & 0.229 ± 0.092 & 0.302 ± 0.089 \\ 
 & IP-E & 0.715 ± 0.107 & 0.261 ± 0.059$^*$ & 0.430 ± 0.278 & 0.169 ± 0.061$^*$ & 0.228 ± 0.048 \\ 
 & IP-A & 0.749 ± 0.116$^*$ & 0.339 ± 0.057$^*$ & 0.530 ± 0.261 & 0.177 ± 0.076$^*$ & 0.247 ± 0.062 \\ 
\bottomrule					
\end{tabular}
\end{table}

\begin{figure}[t]
     \centering

     \begin{subfigure}[b]{0.33\textwidth}
         \centering
         \includegraphics[width=\textwidth,trim=0 0 0 0, clip]{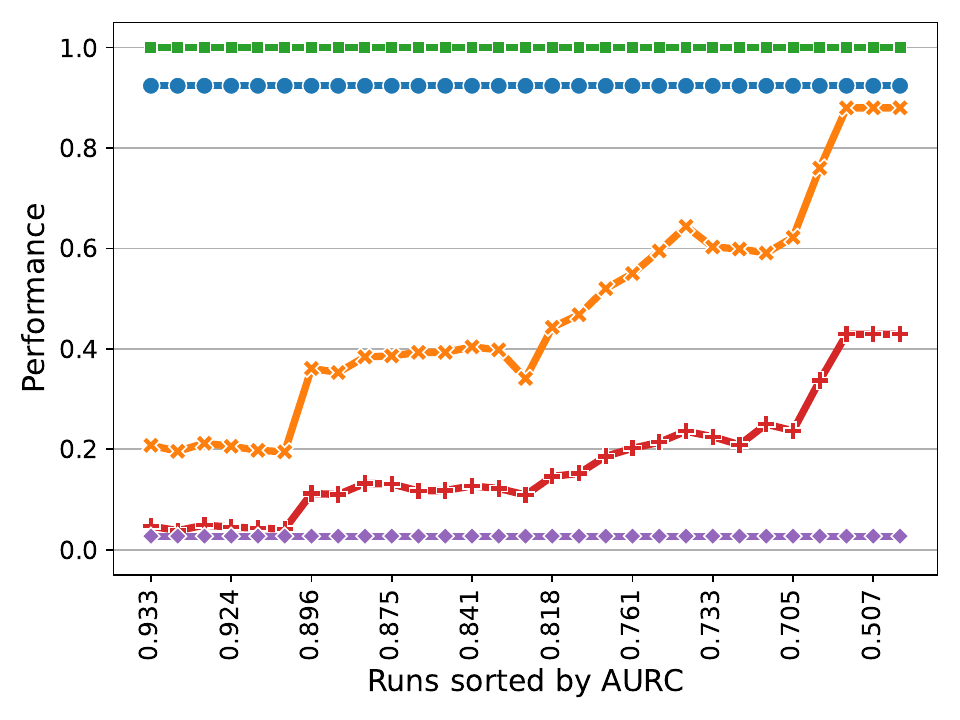}
         \caption{CLEF2017: Oracle}
         \label{fig:clef2017_oracle_0.9}
     \end{subfigure}
     \hfill
     \begin{subfigure}[b]{0.33\textwidth}
         \centering
         \includegraphics[width=\textwidth, trim=0 0 0 0, clip]{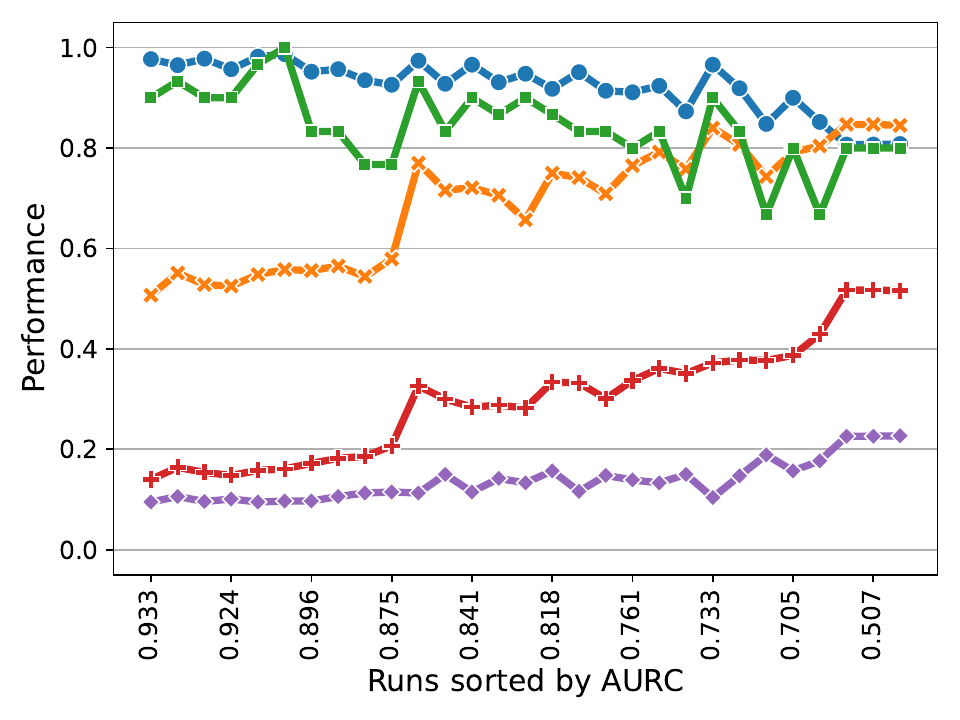}
         \caption{CLEF2017: IP-P}
         \label{fig:clef2017_ip_p_0.9}
     \end{subfigure}
     \hfill
     \begin{subfigure}[b]{0.33\textwidth}
         \centering
         \includegraphics[width=\textwidth, trim=0 0 0 0, clip]{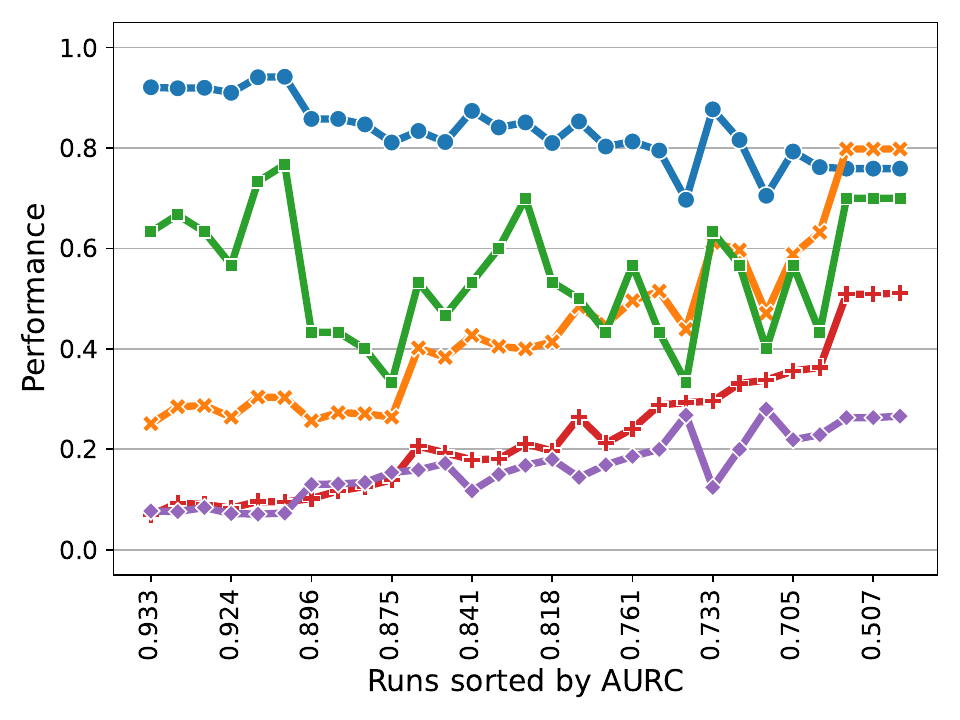}
         \caption{CLEF 2017: IP-E}
         \label{fig:clef2017_ip_e_0.9}
     \end{subfigure}

     \begin{subfigure}[b]{0.33\textwidth}
         \centering
         \includegraphics[width=\textwidth, trim=0 0 0 0, clip]{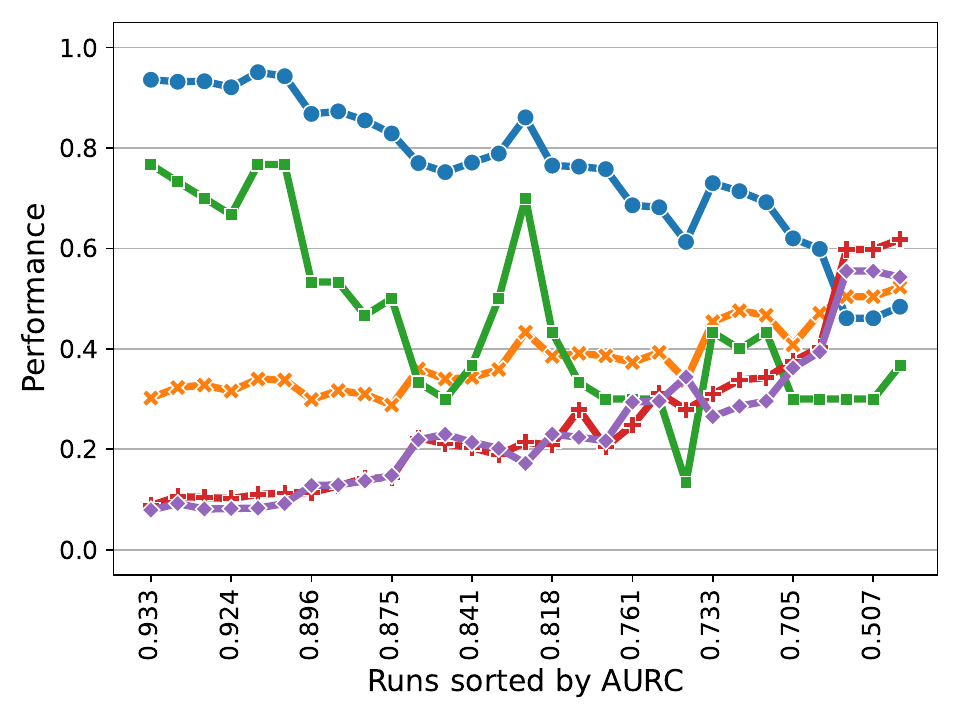}
         \caption{CLEF 2017: IP-A}
         \label{fig:clef2017_ip_a_0.9}
     \end{subfigure}
     \begin{subfigure}[b]{0.33\textwidth}
         \centering
         \includegraphics[width=\textwidth, trim=0 0 0 0, clip]{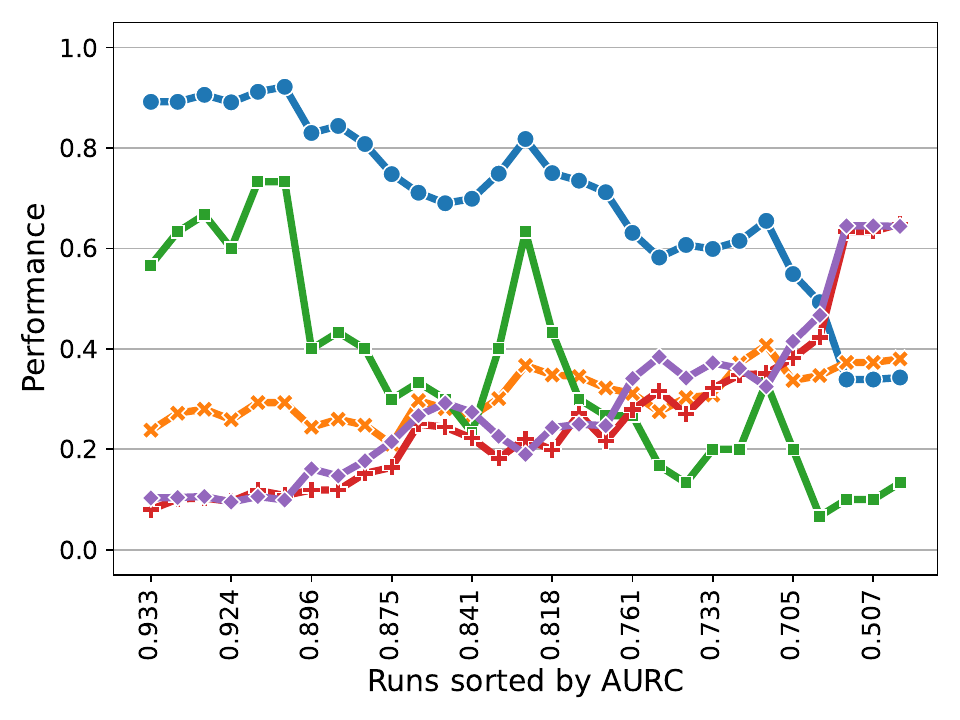}
         \caption{CLEF 2017: IP-H}
         \label{fig:clef2017_ip_h_0.9}
     \end{subfigure}
    \hspace{1.25em}
     \begin{subfigure}[b]{0.30\textwidth}
         \centering
         \includegraphics[width=0.77\textwidth, trim=30 0 0 0, clip]{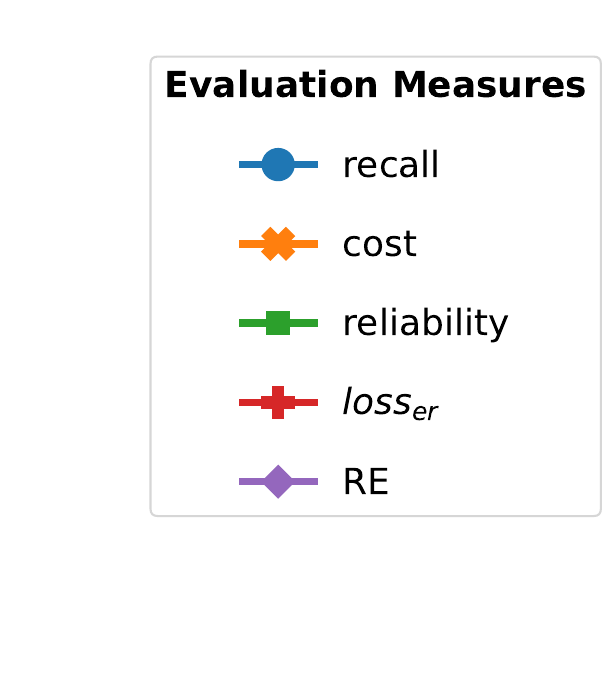}
         \label{fig:clef2017_legend}
     \end{subfigure}

     \begin{subfigure}[b]{0.33\textwidth}
         \centering
         \includegraphics[width=\textwidth, trim=0 0 0 0, clip]{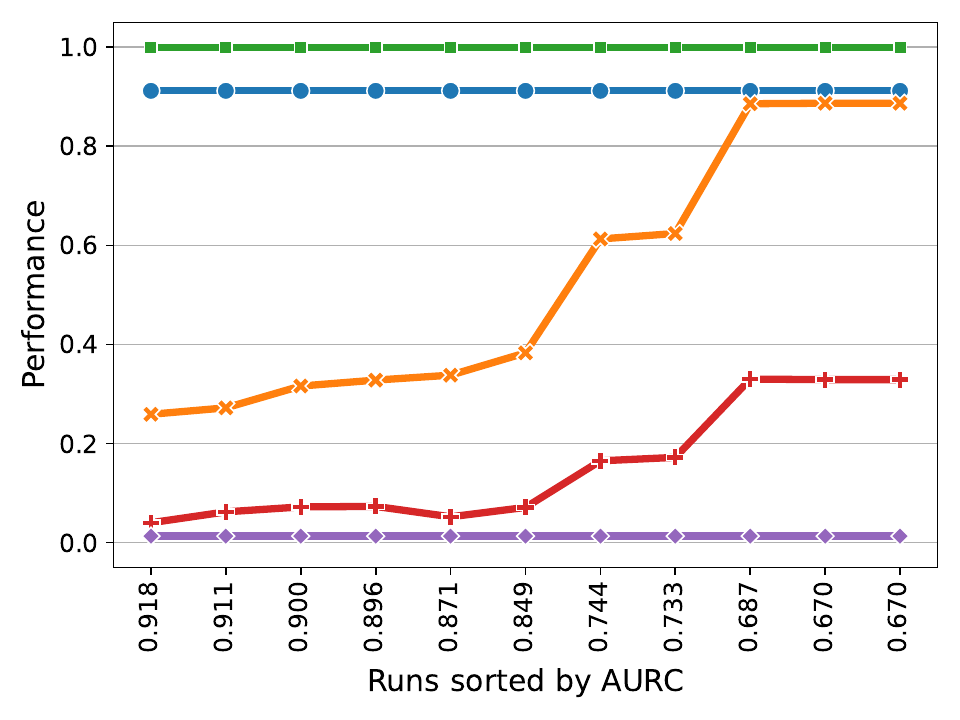}
         \caption{CLEF 2018: Oracle}
         \label{fig:clef2018_oracle_0.9}
     \end{subfigure}
     \hfill
     \begin{subfigure}[b]{0.33\textwidth}
         \centering
         \includegraphics[width=\textwidth, trim=0 0 0 0, clip]{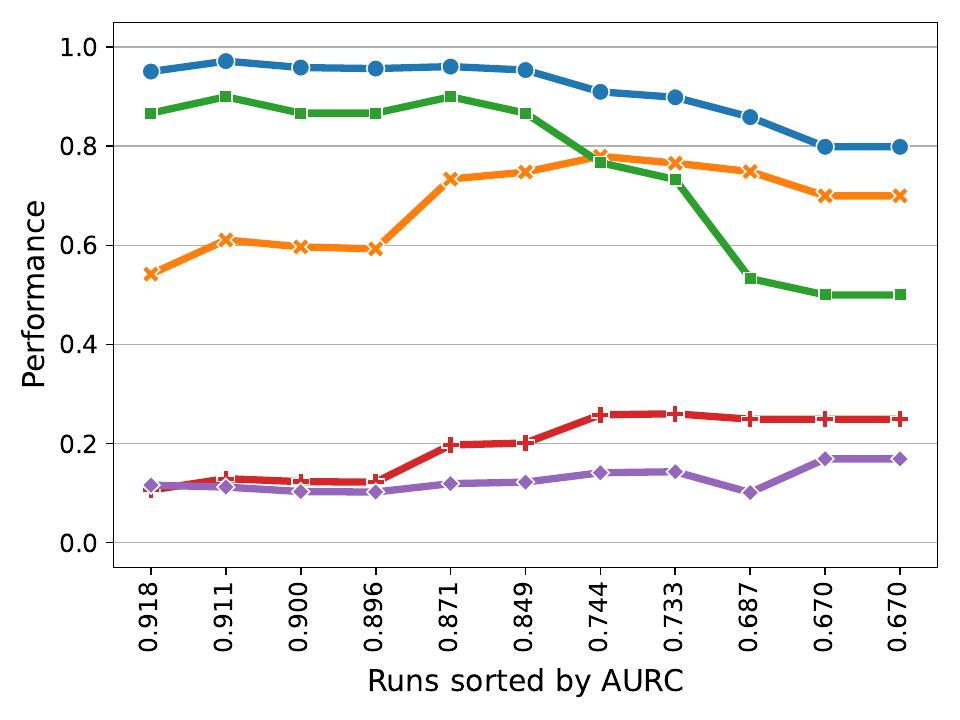}
         \caption{CLEF 2018: IP-P}
         \label{fig:clef2018_ip_p_0.9}
     \end{subfigure}
     \hfill
     \begin{subfigure}[b]{0.33\textwidth}
         \centering
         \includegraphics[width=\textwidth, trim=0 0 0 0, clip]{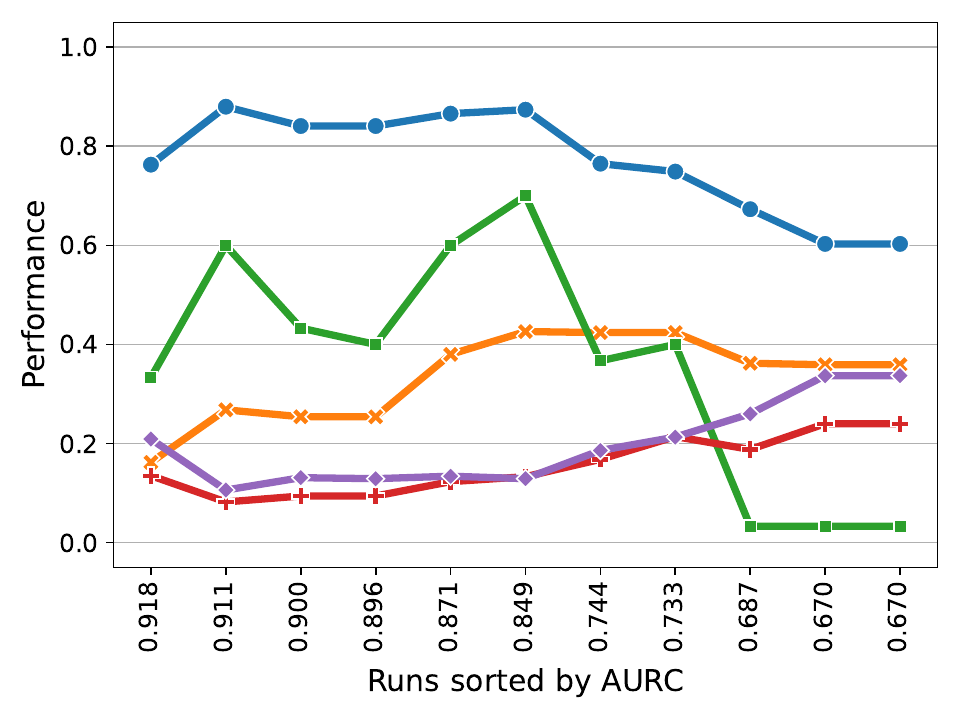}
         \caption{CLEF 2018: IP-E}
         \label{fig:clef2018_ip_e_0.9}
     \end{subfigure}
     \hfill
    \\     
     \begin{subfigure}[b]{0.33\textwidth}
         \centering
         \includegraphics[width=\textwidth, trim=0 0 0 0, clip]{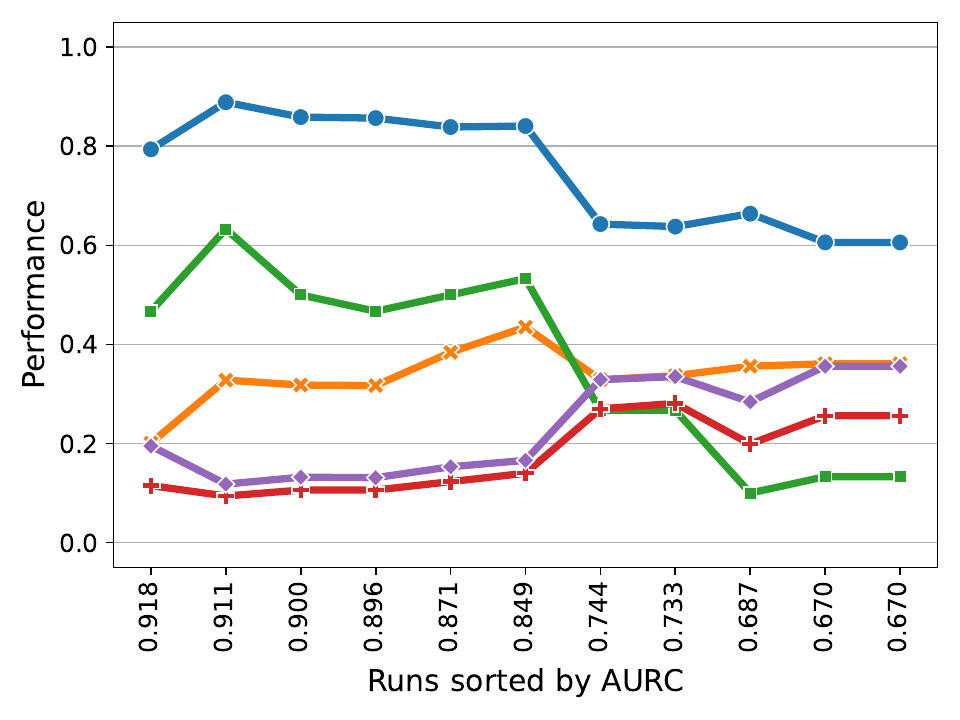}
         \caption{CLEF 2018: IP-A}
         \label{fig:clef2018_ip_a_0.9}
     \end{subfigure}
     \begin{subfigure}[b]{0.33\textwidth}
         \centering
         \includegraphics[width=\textwidth, trim=0 0 0 0, clip]{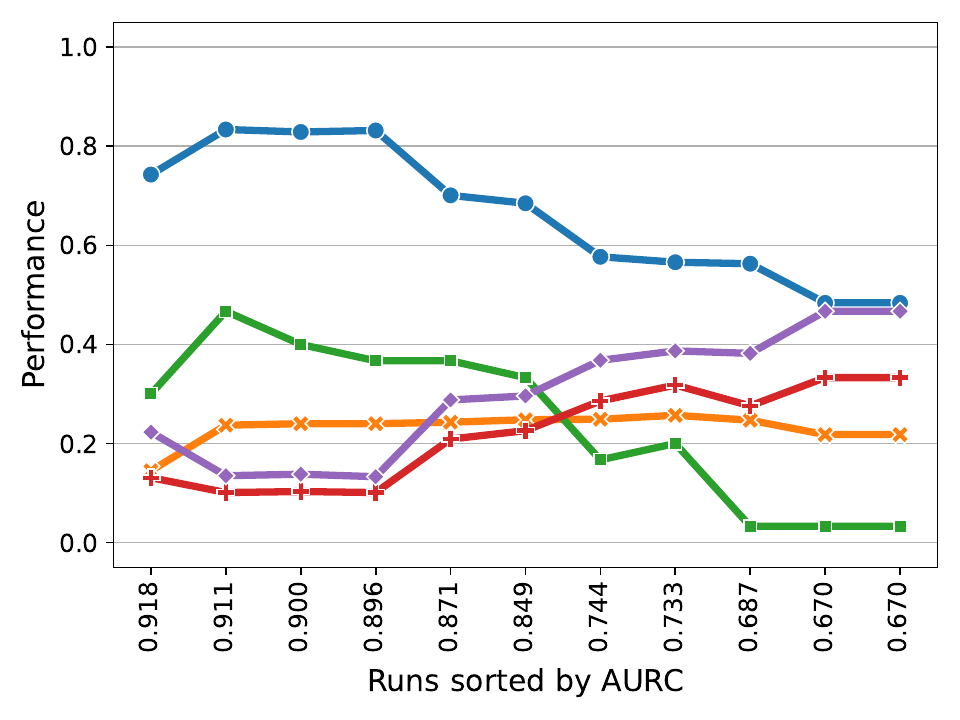}
         \caption{CLEF 2018: IP-H}
         \label{fig:clef2018_ip_h_0.9}
     \end{subfigure}    
     \hfill
    \hspace{1.25em}
     \begin{subfigure}[b]{0.30\textwidth}
         \centering
         \includegraphics[width=\textwidth, trim=0 30 0 250, clip]{figs/clef2018_multiple_runs_tidy_accepted/evaluation_measures_legend_plain.pdf}
         \label{fig:clef2017_legend}
     \end{subfigure}

        \caption{Variation in performance for five metrics for approaches applied to alternative rankings of the CLEF 2017 and 2018 collections with target recall of 0.9. Rankings are sorted by AURC score (descending from left to right).}  
        \label{fig:multiple_runs_clef1718}
\end{figure}

These results are explored in further detail in Figure \ref{fig:multiple_runs_clef1718} which shows the results for each metric for each ranking. These results are sorted by ranking effectiveness which is assessed using the {\tt tar\_eval}'s {\tt norm\_area} metric, the area under the cumulative recall curve normalised by the optimal area. Values reported by Kanoulas et al. \cite{kanoulas2017clef} were used for the CLEF 2017 collection and computed using {\tt tar\_eval} for the CLEF 2018 collection. 

Overall, the reliability is higher with more effective rankings, and it drops sharply with less effective rankings (towards the right of the graph). The drop in reliability is more pronounced for the hyperbolic rate function, which has very low reliability for the less effective rankings (see Figures \ref{fig:clef2017_ip_h_0.9} and \ref{fig:clef2018_ip_h_0.9}).
The decrease in reliability is least pronounced for the power law rate function (see Figures \ref{fig:clef2017_ip_p_0.9}  and \ref{fig:clef2018_ip_p_0.9})  which achieves a reliability of above 80\% for the majority of rankings for the CLEF 2017 dataset and only drops below this value for the least effective of the CLEF 2018 rankings. This suggests that the power law may be a good choice of rate function when robustness is a significant concern or when the ranking effectiveness is unknown, although the number of documents that will need to be examined is likely to be higher than when using other rate functions.

\subsection{Uniform vs Dynamic Batch Sizes}\label{sec:batches}

Our approach analyses documents in batches and only considers stopping points that occur at the end of a batch. In the experiments described so far the batches were uniformly sized as a percentage of the number of documents in the collection. 
An advantage of this approach is that it allows performance to be evaluated using a wide range of different rankings. However, the most successful TAR methods are based on Active Learning approaches, e.g. \cite{cormack2016scalability,li2020stop}, that also analyse documents in batches which determine the number of documents to be manually screened before re-training the classifier. These batches generally increase in size, for example in AutoTAR \cite{cormack2016scalability} the batch size, $B$, is initially set to 1 and then increased by $\left\lceil \frac{B}{10} \right\rceil$ each iteration. Our approach can be naturally adapted to this scenario by altering when the point process is applied to match the batches used by the CAL process. The next experiment explores the effect of making this change.

\begin{table}[!t]
\centering
\caption{IP-H Model Performance Following AutoTAR Batches. Figures in brackets indicate difference between using uniform batch sizes.}\label{tab:CAL_AutoTAR_batches} 
\begin{tabular}{c|ccccc}
\toprule
Dataset  & recall         & cost           & reliability    & loss\textsubscript{er}       & RE       \\					
\midrule
\multicolumn{6}{c}{Target Recall 1.0}                                                         \\
\midrule
CLEF2017 & 0.932 (-0.038) & 0.132 (-0.081) & 0.833 (-0.067) & 0.038 (0.001)  & 0.068 (0.038)  \\
CLEF2018 & 0.917 (-0.039) & 0.153 (-0.059) & 0.733 (-0.067) & 0.055 (0.020)  & 0.083 (0.039)  \\
CLEF2019 & 0.928 (-0.060) & 0.151 (-0.115) & 0.806 (-0.129) & 0.053 (0.004)  & 0.072 (0.060)  \\
Legal    & 0.881 (0.011)  & 0.228 (0.003)  & 0.500 (0.500)  & 0.029 (-0.005) & 0.119 (-0.011) \\
TR       & 1.000 (0.000)  & 0.043 (-0.003) & 0.765 (0.000)  & 0.000 (0.000)  & 0.000 (0.000)  \\
\midrule
\multicolumn{6}{c}{Target Recall 0.9}                                                         \\
\midrule
CLEF2017 & 0.920 (-0.035) & 0.087 (-0.060) & 0.800 (-0.100) & 0.035 (-0.001) & 0.140 (0.017)  \\
CLEF2018 & 0.880 (-0.061) & 0.076 (-0.065) & 0.700 (-0.133) & 0.063 (0.033)  & 0.153 (0.038)  \\
CLEF2019 & 0.921 (-0.063) & 0.110 (-0.104) & 0.839 (-0.129) & 0.048 (0.005)  & 0.145 (0.035)  \\
Legal    & 0.833 (-0.013) & 0.030 (-0.008) & 0.500 (0.000)  & 0.033 (-0.002) & 0.079 (-0.038) \\
TR       & 0.999 (0.000)  & 0.032 (-0.003) & 1.000 (0.000)  & 0.000 (0.000)  & 0.110 (0.000)  \\
\midrule
\multicolumn{6}{c}{Target Recall 0.8}                                                         \\
\midrule
CLEF2017 & 0.906 (-0.043) & 0.079 (-0.058) & 0.833 (-0.067) & 0.038 (0.001)  & 0.224 (-0.007) \\
CLEF2018 & 0.871 (-0.065) & 0.070 (-0.064) & 0.867 (-0.066) & 0.064 (0.033)  & 0.239 (0.024)  \\
CLEF2019 & 0.916 (-0.066) & 0.102 (-0.107) & 0.871 (-0.097) & 0.047 (0.004)  & 0.241 (0.003)  \\
Legal    & 0.813 (0.020)  & 0.027 (0.002)  & 0.500 (0.000)  & 0.038 (-0.007) & 0.063 (-0.002) \\
TR       & 0.999 (0.000)  & 0.032 (-0.003) & 1.000 (0.000)  & 0.000 (0.000)  & 0.249 (0.000)  \\
\bottomrule					
\end{tabular}
\end{table}

Our approach was adapted to follow the same batches used by AutoTAR  \cite{cormack2016scalability}. 
The IP-H model was used with the model applied to the ranking produced by AutoStop, which also follows AutoTAR batches. Results are shown in Table \ref{tab:CAL_AutoTAR_batches} where the figures in brackets indicate the difference between the corresponding scores obtained using uniform batch sizes. 
The overall results show that changing to dynamic batches tends to produce a small decrease in recall, cost and reliability. The drop in reliability is fairly substantial in some cases but the corresponding differences in recall indicate that the number of relevant documents identified was similar.
It is worth noting that, although the batch sizes used by AutoTAR are well suited for CAL frameworks, they are not ideal for the stopping problem. AutoTAR batches are independent of the collection size and start very small then gradually increase in size but, since they dictate the set of candidate stopping points, this reduces the number of places at which the algorithm can stop later in the ranking. 
This mismatch is the likely reason for the reduction in performance when the AutoTAR batches are used. 

\section{Conclusion}

This paper explored the problem of deciding when to stop examining documents in a ranked list so that a specified portion of the relevant ones has been identified. The proposed approach is based on point processes, which can be used to model the occurrence of random events over time, and are applied to model the rate at which relevant documents are encountered in a ranked list of documents. Two point processes (Inhomogeneous Poisson and Cox Processes) and four rate functions (exponential, power law, AP Prior and hyperbolic decline) were compared and evaluated using five data sets (CLEF  Technology-Assisted Review in Empirical Medicine 2017-9, TREC Total Recall and TREC Legal). Experiments demonstrated that in the majority of cases, the proposed approach is able to identify a stopping point that achieves the target recall without requiring an excessive number of documents to be examined. It also performed well in comparison to a range of alternative stopping methods. Two of these alternative methods, the generalisation of the target and QBCB methods, were more likely to achieve the target recall than our proposed approach but at the cost of requiring more documents to be examined. Results also showed that employing different rate functions varied the behaviour of the proposed approach with hyperbolic decline leading to a balance between reaching target recall and the number of documents examined. Using the power law as a rate function was more reliable but required more documents to be examined. Results also showed that there was little difference in performance between the Inhomogeneous Poisson Process and the more computationally expensive Cox Process. Further experiments were carried out using a range of rankings of varying effectiveness. They demonstrated that the number of documents that need to be examined to reach a particular recall increases for less effective rankings. They also showed that the proposed approach remains reliable across a wide range of rankings when the power law rate function is used while the reliability tends to drop (often substantially) when other rate functions are used. 

\subsection{Discussion and Future Work} 

This work has demonstrated the importance of the ranking in stopping algorithm effectiveness. While this relationship is perhaps unsurprising it has, to the best of our knowledge, not previously been demonstrated empirically. This highlights a more general issue with the evaluation of stopping algorithms since previous approaches have been evaluated different rankings (not all of which are generally available) with each algorithm invariably being evaluated using a single ranking. The community would therefore benefit from access to a common set of retrieval problems and rankings against which stopping algorithms could be evaluated. These rankings should include those generated by neural methods, which have recently shown promise for high-recall tasks \cite{yang2022goldilocks,wang2022neural}.

The proposed approach models the number of relevant documents remaining using a Poisson distribution which has the highly restrictive assumption that the variance equals the mean. This can be problematic, particularly in situations where the estimated number of documents is high since the variance will also be high. Future work will explore ways to mitigate this limitation. 

Another potential avenue for future work would be to integrate a classifier into the stopping algorithm, similar to \citet{Yang2021heuristic} and \citet{yu2019fast2} (see Section \ref{sec:no_rel}). The classifier could be trained using the relevance judgements available from the trained part of the ranking and then applied to the unobserved part. Its output would provide information about the likelihood of those documents being relevant that could then be used by the point process to improve the estimate of the number of relevant documents remaining.

In common with previous work on stopping methods for TAR, the work described here focuses on the problem of achieving a specified target recall, i.e. identifying a set portion of the relevant documents. However, recall does not take account of the effort required to identify relevant documents which can vary considerably depending on their prevalence. An alternative approach to developing stopping algorithms could be to continue until the effort required becomes excessive. A potential method for assessing effort is available from the field of systematic reviews where the number needed to read metric measures the number of documents that need to be examined in order to find a single relevant one, i.e. reciprocal of precision \cite{cooper2018systematic}. In addition, the recall achieved is often less important than whether an information need has been met. For example, in medicine Diagnostic Test Accuracy systematic reviews aim to quantify the effectiveness of medical tests (in terms of specificity and sensitivity). \citet{norman2019measuring} developed stopping criteria based on the reliability (or variance) of these estimates, rather than when a specified proportion of the evidence has been identified. Another possible route for future work would be to extend the approaches described in this paper to estimate the amount of information remaining, and the possibility that it would alter the conclusions that had been drawn from the documents examined so far.

Finally, work on stopping methods, including the approach presented here, relies on the assumption that relevance judgements provided by assessors are reliable and consistent. However, it has long been known that this is not the case, e.g. \cite{katter1968influence,scholer2011quantifying,belur2021interrater}, which could have a significant effect on stopping algorithms since their decisions may be based on relatively small numbers of relevant documents. Exploring the relationship between relevance judgement consistency and the effectiveness of stopping algorithms represents an interesting direction for future work.

\section*{Acknowledgements}

The authors are grateful to Alison Sneyd for her contributions to earlier versions of the approach described here \cite{Sneyd19,Sneyd21}.

\bibliographystyle{apalike}
\bibliography{references}

\appendix

\section{Generalisation of Target Method}\label{appendix:target}

The original description of the target method \cite{cormack2016engineering} includes a proof to demonstrate that recall of 0.7 can be guaranteed in 95\% of cases by setting the target value, $t$, to 10. That proof is now generalised for arbitrary target recalls ($\ell$) and confidence values ($c$). 

The target method randomly draws documents from the collection until $t$ relevant ones are identified. All documents until the highest ranked of the randomly chosen relevant documents are then returned. We aim to find a value for $t$ such that a recall of (at least) $\ell$ is achieved with probability $c$. 

Consider a ranked list of documents, $D$. The rank of a document $d \in D$ is denoted by $rank(d)$ and $rel(d)$ is a binary function that is true if and only if $d$ is relevant. 

The set $T$ is constructed by randomly drawing documents from $D$ until it contains $t$ relevant documents. Now consider the set $U$ constructed by choosing all documents in $D$ with rank no greater than that of the highest ranked relevant document in $T$. More formally, let $T_{rel}$ be the subset of $T$ that are relevant, i.e. $T_{rel} = \{ d \in T: rel(d)\}$ then the highest ranked relevant document, $T_{last\_rel}$, is $arg\,max_{d \in T_{rel}} rank(d)$. Then, $U  = \{ d \in D: rel(d) \land rank(d) \leq rank(T_{last\_rel}) \}$. 

$U$ is the set of relevant documents returned by the target method. Let $|U| = u$ and assume that the collection contains $r$ relevant documents then the recall will be  $\frac{u}{r}$.

The desired recall, $\ell$, is achieved with a desired probability, $c$, if 
\begin{displaymath}
P \left( \frac{u}{r} \geq \ell \right) \geq c
\end{displaymath}

which holds if

\begin{displaymath}
P \left( \frac{r - u}{r} \geq 1 - \ell \right) \leq 1 - c \,.
\end{displaymath}

Now consider the problem of finding a value $\alpha$ such that 
\begin{equation}\label{eq:proof}
P \left( \frac{r - u}{r} \geq \alpha \right) \leq 1 - c 
\end{equation}
which can be rewritten as 
\begin{displaymath}
P \left( r - u \geq \alpha r \right) \leq 1 - c .
\end{displaymath}

For this to be true there must be $\alpha r$ relevant documents not included in $U$. This probability of which can be estimated as\footnote{This proof follows \citet{cormack2016engineering} in modelling this probability using a Binomial distribution, i.e. assuming sampling with replacement. Sampling without replacement would arguably be more appropriate since it is unlikely that a document would be examined for relevance more than once but we choose to follow the previous approach as closely as possible.}
\begin{displaymath}
\left( 1 - \frac{t}{r} \right)^{\alpha r} = 1 - c \,,
\end{displaymath}

i.e. 
\begin{displaymath}
\alpha = \frac{\log(1 -c)}{r \log(1 - \frac{t}{r})} \,.
\end{displaymath}

Now consider the situation where $r$ is large, 
\begin{displaymath}
\lim_{r \to \infty} \frac{\log(1 -c)}{r \log(1 - \frac{t}{r})} = - \frac{\log(1 -c)}{t} \;\;\textrm{(by L'H\^{o}pital's Rule).} 
\end{displaymath}

Combining this with \ref{eq:proof} produces 
\begin{displaymath}
t = - \frac{\log(1 -c)}{1 - \ell} \,.
\end{displaymath}

\newpage
\section{Detailed Results for Comparison with Alternative Approaches}\label{sec:detailed_baseline_results}

\begin{table}[!th]
\caption{Baseline comparison for target recall of 1.0. Results for TM, Knee, SCAL, SD-training, SD-sampling and AutoStop from \citet{li2020stop}.}\label{tab:0.99recall_all}
\resizebox*{!}{0.89\textheight}{%
\begin{tabular}{l|l|rrrrr}
\toprule
Dataset &  &  recall &   cost &  reliability &  loss\textsubscript{er}  &  RE \\
\midrule
& OR            &  1.000 ± 0.000 &  0.133 ± 0.194 &  1.000 &   0.005 ± 0.007 &  0.000 ± 0.000 \\
& TM            &  0.978 ± 0.036 &  0.614 ± 0.258 &  0.567 &   0.292 ± 0.349 &  0.022 ± 0.036 \\
& TM-adapted    &  1.000 ± 0.001 &  0.993 ± 0.028 &  1.000 &  0.468 ± 0.293 &  0.010 ± 0.001 \\
& QBCB          &  0.985 ± 0.027 &  0.656 ± 0.357 &  0.700 &  0.366 ± 0.371 &  0.018 ± 0.021 \\
& Knee          &  0.998 ± 0.006 &  0.291 ± 0.194 &  0.833 &   0.041 ± 0.081 &  0.002 ± 0.006 \\
CLEF2017 & SCAL &  0.984 ± 0.038 &  0.659 ± 0.281 &  0.700 &   0.253 ± 0.265 &  0.016 ± 0.038 \\
& SD-training   &  0.999 ± 0.003 &  0.997 ± 0.007 &  0.967 &   0.466 ± 0.292 &  0.010 ± 0.001 \\
& SD-sampling   &  0.973 ± 0.042 &  0.762 ± 0.283 &  0.533 &   0.311 ± 0.320 &  0.027 ± 0.036 \\
& AutoStop      &  0.999 ± 0.008 &  0.625 ± 0.190 &  0.967 &   0.161 ± 0.115 &  0.001 ± 0.008 \\
& IP-H          &  0.970 ± 0.095 &  0.213 ± 0.187 &  0.900 &   0.037 ± 0.043 &  0.030 ± 0.095 \\
				
\midrule
& OR            &  1.000 ± 0.000 &  0.161 ± 0.207 &  1.000 &   0.007 ± 0.013 &  0.000 ± 0.000 \\
& TM            &  0.983 ± 0.022 &  0.594 ± 0.289 &  0.500 &   0.251 ± 0.310 &  0.017 ± 0.022 \\
& TM-adapted    &  1.000 ± 0.000 &  0.990 ± 0.050 &  1.000 &  0.396 ± 0.279 &  0.010 ± 0.000 \\
& QBCB          &  0.974 ± 0.059 &  0.590 ± 0.380 &  0.633 &  0.302 ± 0.336 &  0.028 ± 0.054 \\
& Knee          &  0.994 ± 0.013 &  0.328 ± 0.258 &  0.700 &   0.047 ± 0.080 &  0.006 ± 0.013 \\
CLEF2018 & SCAL &  0.982 ± 0.039 &  0.657 ± 0.284 &  0.600 &   0.233 ± 0.254 &  0.018 ± 0.039 \\
& SD-training   &  1.000 ± 0.000 &  0.996 ± 0.008 &  1.000 &   0.394 ± 0.277 &  0.010 ± 0.000 \\
& SD-sampling   &  0.964 ± 0.093 &  0.688 ± 0.270 &  0.533 &   0.180 ± 0.170 &  0.036 ± 0.090 \\
& AutoStop      &  0.992 ± 0.044 &  0.662 ± 0.182 &  0.967 &   0.163 ± 0.136 &  0.008 ± 0.044 \\
& IP-H          &  0.956 ± 0.127 &  0.212 ± 0.168 &  0.800 &   0.035 ± 0.062 &  0.044 ± 0.127 \\
				
\midrule
& OR            &  1.000 ± 0.000 &  0.116 ± 0.123 &  1.000 &   0.009 ± 0.017 &  0.000 ± 0.000 \\
& TM            &  0.985 ± 0.025 &  0.753 ± 0.275 &  0.645 &   0.428 ± 0.350 &  0.015 ± 0.025 \\
& TM-adapted    &  1.000 ± 0.000 &  1.000 ± 0.001 &  1.000 &  0.552 ± 0.261 &  0.010   ±   0.000 \\
& QBCB          &  0.983 ± 0.036 &  0.742 ± 0.371 &  0.710 &  0.460 ± 0.359 &  0.021 ± 0.030 \\
& Knee          &  0.998 ± 0.009 &  0.420 ± 0.311 &  0.968 &   0.122 ± 0.160 &  0.002 ± 0.009 \\
CLEF2019 & SCAL &  0.993 ± 0.022 &  0.808 ± 0.219 &  0.839 &   0.410 ± 0.288 &  0.007 ± 0.022 \\
& SD-training   &  0.999 ± 0.004 &  0.992 ± 0.016 &  0.968 &   0.545 ± 0.262 &  0.010 ± 0.001 \\
& SD-sampling   &  0.974 ± 0.070 &  0.767 ± 0.249 &  0.623 &   0.306 ± 0.273 &  0.028 ± 0.066 \\
& AutoStop      &  1.000 ± 0.000 &  0.651 ± 0.198 &  1.000 &   0.223 ± 0.148 &  0.000 ± 0.000 \\
& IP-H          &  0.988 ± 0.061 &  0.266 ± 0.201 &  0.935 &   0.049 ± 0.062 &  0.012 ± 0.061 \\
				
\midrule
& OR            &  1.000 ± 0.000 &  0.043 ± 0.108 &  1.000 &   0.000 ± 0.000 &  0.000 ± 0.000 \\
& TM            &  0.944 ± 0.063 &  0.120 ± 0.150 &  0.147 &   0.024 ± 0.068 &  0.056 ± 0.063 \\
& TM-adapted    &  0.999 ± 0.001 &  0.718 ± 0.303 &  1.000 &  0.118 ± 0.175 &  0.009 ± 0.001 \\
& QBCB          &  0.986 ± 0.019 &  0.142 ± 0.279 &  0.559 &  0.054 ± 0.173 &  0.013 ± 0.014 \\
& Knee          &  0.960 ± 0.056 &  0.016 ± 0.041 &  0.088 &   0.005 ± 0.016 &  0.040 ± 0.056 \\
TR & SCAL       &  0.991 ± 0.168 &  0.146 ± 0.371 &  0.147 &   0.079 ± 0.211 &  0.081 ± 0.168 \\
& SD-training   &  1.000 ± 0.000 &  1.000 ± 0.000 &  1.000 &   0.122 ± 0.173 &  0.010 ± 0.000 \\
& SD-sampling   &  0.988 ± 0.066 &  0.944 ± 0.225 &  0.941 &   0.106 ± 0.142 &  0.022 ± 0.063 \\
& AutoStop      &  1.000 ± 0.000 &  0.779 ± 0.148 &  0.941 &   0.100 ± 0.160 &  0.000 ± 0.000 \\
& IP-H          &  1.000 ± 0.001 &  0.046 ± 0.064 &  0.765 &   0.000 ± 0.000 &  0.000 ± 0.001 \\

\midrule
& OR            &  1.000 ± 0.000 &  0.326 ± 0.146 &  1.000 &   0.001 ± 0.001 &  0.000 ± 0.000 \\
& TM            &  0.953 ± 0.028 &  0.147 ± 0.008 &  0.000 &   0.003 ± 0.003 &  0.047 ± 0.028 \\
& TM-adapted    &  1.000 ± 0.001 &  0.660 ± 0.004 &  1.000 &  0.004 ± 0.001 &  0.010 ± 0.001 \\
& QBCB          &  0.979 ± 0.013 &  0.194 ± 0.000 &  0.000 &  0.001 ± 0.001 &  0.011 ± 0.013 \\
& Knee          &  0.966 ± 0.048 &  0.287 ± 0.272 &  0.500 &   0.004 ± 0.002 &  0.034 ± 0.048 \\
Legal & SCAL    &  0.039 ± 0.031 &  0.005 ± 0.001 &  0.000 &   0.924 ± 0.060 &  0.961 ± 0.031 \\
& SD-training   &  1.000 ± 0.000 &  1.000 ± 0.000 &  1.000 &   0.007 ± 0.002 &  0.101 ± 0.000 \\
& SD-sampling   &  1.000 ± 0.000 &  1.000 ± 0.000 &  1.000 &   0.007 ± 0.002 &  0.010 ± 0.000 \\
& AutopStop     &  0.996 ± 0.001 &  0.833 ± 0.025 &  0.000 &   0.005 ± 0.001 &  0.004 ± 0.001 \\
& IP-H          &  0.870 ± 0.182 &  0.225 ± 0.283 &  0.000 &   0.034 ± 0.047 &  0.130 ± 0.182 \\
				
\bottomrule
\end{tabular}
}
\end{table}

\begin{table}[!h]
\caption{Baseline comparison for target recall of 0.9. Results for TM, Knee, SCAL, SD-training, SD-sampling and AutoStop from \citet{li2020stop}.}\label{tab:0.9recall_all}
\begin{tabular}{l|l|rrrrr}
\toprule
Dataset & model 0.9 &  recall &   cost &  reliability &  loss\textsubscript{er} &  RE \\
\midrule
& OR            & 0.923 ± 0.057 & 0.061 ± 0.062 & 1.000 & 0.009 ± 0.006 & 0.026 ± 0.039 \\
& TM-adapted    & 0.990 ± 0.016 &  0.736 ± 0.240 &  1.000 &  0.366 ± 0.349 &  0.100 ± 0.017 \\
& QBCB          & 0.975 ± 0.039 &  0.641 ± 0.374 &  0.933 &  0.365 ± 0.371 &  0.088 ± 0.034 \\
& SCAL          & 0.914 ± 0.075 & 0.496 ± 0.244 & 0.667 & 0.168 ± 0.209 & 0.072 ± 0.042 \\
CLEF2017 & SD-training   & 0.955 ± 0.057 & 0.691 ± 0.054 & 0.833 & 0.233 ± 0.148 & 0.080 ± 0.034 \\
& SD-sampling   & 0.902 ± 0.083 & 0.506 ± 0.277 & 0.567 & 0.192 ± 0.278 & 0.071 ± 0.057 \\
& AutoStop      & 0.884 ± 0.088 & 0.421 ± 0.097 & 0.500 & 0.097 ± 0.065 & 0.069 ± 0.070 \\
& IP-H          & 0.955 ± 0.119 & 0.147 ± 0.114 & 0.900 & 0.036 ± 0.055 & 0.123 ± 0.076 \\
				
\midrule
& OR            & 0.912 ± 0.019 & 0.067 ± 0.064 & 1.000 & 0.010 ± 0.004 & 0.013 ± 0.021 \\
& TM-adapted    & 0.992 ± 0.012 &  0.729 ± 0.254 &  1.000 &  0.312 ± 0.312 &  0.102 ± 0.013 \\
& QBCB          & 0.968 ± 0.060 &  0.559 ± 0.401 &  0.900 &  0.300 ± 0.338 &  0.094 ± 0.033 \\
& SCAL          & 0.902 ± 0.087 & 0.493 ± 0.241 & 0.667 & 0.154 ± 0.168 & 0.074 ± 0.060 \\
CLEF2018 & SD-training   & 0.972 ± 0.033 & 0.701 ± 0.038 & 0.967 & 0.196 ± 0.138 & 0.082 ± 0.030 \\
& SD-sampling   & 0.855 ± 0.108 & 0.379 ± 0.217 & 0.367 & 0.077 ± 0.074 & 0.080 ± 0.102 \\
& AutoStop      & 0.892 ± 0.075 & 0.441 ± 0.103 & 0.600 & 0.098 ± 0.078 & 0.046 ± 0.071 \\
& IP-H          & 0.941 ± 0.125 & 0.141 ± 0.145 & 0.833 & 0.030 ± 0.063 & 0.115 ± 0.088 \\
				
\midrule
& OR            & 0.929 ± 0.037 & 0.071 ± 0.082 & 1.000 & 0.011 ± 0.013 & 0.033 ± 0.041 \\
& TM-adapted    & 0.994 ± 0.013 &  0.854 ± 0.205 &  1.000 &  0.478 ± 0.329 &  0.104 ± 0.015 \\
& QBCB          & 0.980 ± 0.036 &  0.705 ± 0.371 &  0.935 &  0.440 ± 0.370 &  0.092 ± 0.031 \\
& SCAL          & 0.893 ± 0.104 & 0.621 ± 0.206 & 0.516 & 0.271 ± 0.198 & 0.082 ± 0.082 \\
CLEF2019 & SD-training   & 0.940 ± 0.100 & 0.713 ± 0.043 & 0.774 & 0.295 ± 0.156 & 0.092 ± 0.075 \\
& SD-sampling   & 0.893 ± 0.095 & 0.517 ± 0.270 & 0.50  & 0.198 ± 0.260 & 0.072 ± 0.077 \\
& AutoStop      & 0.878 ± 0.096 & 0.479 ± 0.129 & 0.387 & 0.159 ± 0.154 & 0.072 ± 0.080 \\
& IP-H          & 0.984 ± 0.061 & 0.214 ± 0.177 & 0.968 & 0.043 ± 0.063 & 0.110 ± 0.032 \\
				
\midrule
& OR            & 0.902 ± 0.004 & 0.005 ± 0.010 & 1.000 & 0.010 ± 0.001 & 0.003 ± 0.004 \\
& TM-adapted    & 0.977 ± 0.029 &  0.216 ± 0.246 &  0.971 &  0.052 ± 0.155 &  0.088 ± 0.025 \\
& QBCB          & 0.972 ± 0.026 &  0.115 ± 0.282 &  0.971 &  0.054 ± 0.173 &  0.081 ± 0.028 \\
& SCAL          & 0.903 ± 0.171 & 0.144 ± 0.318 & 0.647 & 0.083 ± 0.210 & 0.094 ± 0.163 \\
TR & SD-training   & 1.000 ± 0.000 & 1.000 ± 0.000 & 1.000 & 0.122 ± 0.173 & 0.111 ± 0.000 \\
& SD-sampling   & 0.936 ± 0.129 & 0.779 ± 0.407 & 0.794 & 0.102 ± 0.136 & 0.133 ± 0.063 \\
& AutoStop      & 0.953 ± 0.030 & 0.766 ± 0.163 & 0.941 & 0.103 ± 0.158 & 0.062 ± 0.027 \\
& IP-H       & 0.999 ± 0.002 & 0.035 ± 0.024 & 1.000 & 0.000 ± 0.000 & 0.110 ± 0.002 \\
			
\midrule
& OR            & 0.900 ± 0.000 & 0.040 ± 0.011 & 1.000 & 0.010 ± 0.000 & 0.000 ± 0.000 \\
& TM-adapted    & 0.985 ± 0.005 &  0.223 ± 0.012 &  1.000 &  0.001 ± 0.001 &  0.094 ± 0.006 \\
& QBCB          & 0.948 ± 0.012 &  0.129 ± 0.043 &  1.000 &  0.003 ± 0.001 &  0.053 ± 0.013 \\
& SCAL          & 0.039 ± 0.031 & 0.005 ± 0.001 & 0.000 & 0.924 ± 0.060 & 0.957 ± 0.035 \\
Legal & SD-training   & 1.000 ± 0.000 & 1.000 ± 0.000 & 1.000 & 0.007 ± 0.002 & 0.111 ± 0.000 \\
& SD-sampling   & 1.000 ± 0.000 & 1.000 ± 0.000 & 1.000 & 0.007 ± 0.002 & 0.111 ± 0.000 \\
& AutoStop      & 0.803 ± 0.006 & 0.811 ± 0.029 & 0.000 & 0.043 ± 0.004 & 0.108 ± 0.007 \\
& IP-H    & 0.846 ± 0.148 & 0.038 ± 0.018 & 0.500 & 0.035 ± 0.046 & 0.117 ± 0.085 \\
				
\bottomrule
\end{tabular}
\end{table}


\begin{table}[!h]
\caption{Baseline comparison for target recall of 0.8. Results for TM, Knee, SCAL, SD-training, SD-sampling and AutoStop from \citet{li2020stop}.}\label{tab:0.8recall_all}
\begin{tabular}{l|l|rrrrr}
\toprule
Dataset & &  recall &   cost &  reliability &  loss\textsubscript{er} &  RE \\
\midrule
& OR            & 0.834 ± 0.067 & 0.043 ± 0.048 & 1.000 & 0.033 ± 0.014 & 0.042 ± 0.084 \\
& TM-adapted    & 0.969 ± 0.035 &  0.570 ± 0.279 &  1.000 &  0.277 ± 0.357 &  0.212 ± 0.044 \\
& QBCB          & 0.945 ± 0.056 &  0.470 ± 0.418 &  1.000 &  0.300 ± 0.390 &  0.181 ± 0.070 \\
& SCAL          & 0.888 ± 0.089 & 0.451 ± 0.255 & 0.800 & 0.177 ± 0.246 & 0.138 ± 0.072 \\
CLEF2017 & SD-training   & 0.881 ± 0.113 & 0.417 ± 0.068 & 0.767 & 0.109 ± 0.062 & 0.148 ± 0.088 \\
& SD-sampling   & 0.798 ± 0.087 & 0.350 ± 0.270 & 0.433 & 0.170 ± 0.269 & 0.077 ± 0.076 \\
& AutoStop      & 0.787 ± 0.090 & 0.335 ± 0.077 & 0.367 & 0.105 ± 0.056 & 0.088 ± 0.080 \\
& IP-H          & 0.949 ± 0.120 & 0.137 ± 0.113 & 0.900 & 0.037 ± 0.055 & 0.231 ± 0.056 \\
				
\midrule
& OR            & 0.812 ± 0.016 & 0.051 ± 0.054 & 1.000 & 0.037 ± 0.006 & 0.015 ± 0.020 \\
& TM-adapted    & 0.985 ± 0.020 &  0.585 ± 0.281 &  1.000 &  0.234 ± 0.306 &  0.231 ± 0.025 \\
& QBCB          & 0.940 ± 0.074 &  0.518 ± 0.432 &  0.933 &  0.300 ± 0.338 &  0.186 ± 0.065 \\
& SCAL          & 0.862 ± 0.093 & 0.428 ± 0.245 & 0.767 & 0.144 ± 0.162 & 0.119 ± 0.070 \\
CLEF2018 & SD-training   & 0.886 ± 0.094 & 0.414 ± 0.059 & 0.833 & 0.086 ± 0.055 & 0.141 ± 0.074 \\
& SD-sampling   & 0.753 ± 0.137 & 0.258 ± 0.168 & 0.367 & 0.099 ± 0.100 & 0.121 ± 0.134 \\
& AutoStop      & 0.781 ± 0.073 & 0.347 ± 0.089 & 0.467 & 0.104 ± 0.061 & 0.064 ± 0.069 \\
& IP-H          & 0.936 ± 0.124 & 0.134 ± 0.146 & 0.933 & 0.031 ± 0.063 & 0.215 ± 0.079 \\
				
\midrule
& OR            & 0.830 ± 0.050 & 0.057 ± 0.069 & 1.000 & 0.035 ± 0.015 & 0.037 ± 0.062 \\
& TM-adapted    & 0.987 ± 0.021 &  0.753 ± 0.273 &  1.000 &  0.416 ± 0.353 &  0.234 ± 0.026 \\
& QBCB          & 0.962 ± 0.058 &  0.676 ± 0.385 &  0.968 &  0.432 ± 0.372 &  0.204 ± 0.068 \\
& SCAL          & 0.887 ± 0.086 & 0.577 ± 0.245 & 0.903 & 0.261 ± 0.228 & 0.134 ± 0.071 \\
CLEF2019 & SD-training   & 0.826 ± 0.153 & 0.421 ± 0.066 & 0.613 & 0.146 ± 0.085 & 0.155 ± 0.114 \\
& SD-sampling   & 0.787 ± 0.125 & 0.366 ± 0.249 & 0.475 & 0.166 ± 0.217 & 0.111 ± 0.110 \\
& AutoStop      & 0.791 ± 0.121 & 0.397 ± 0.119 & 0.452 & 0.158 ± 0.143 & 0.111 ± 0.100 \\
& IP-H          & 0.982 ± 0.062 & 0.209 ± 0.178 & 0.968 & 0.043 ± 0.063 & 0.238 ± 0.026 \\
				
\midrule
& OR            & 0.802 ± 0.003 & 0.004 ± 0.008 & 1.000 & 0.039 ± 0.001 & 0.003 ± 0.004 \\
& TM-adapted    & 0.946 ± 0.059 &  0.114 ± 0.131 &  0.971 &  0.020 ± 0.051 &  0.183 ± 0.072 \\
& QBCB          & 0.910 ± 0.047 &  0.096 ± 0.286 &  0.971 &  0.063 ± 0.170 &  0.138 ± 0.057 \\
& SCAL          & 0.761 ± 0.288 & 0.107 ± 0.282 & 0.676 & 0.167 ± 0.285 & 0.247 ± 0.263 \\
TR & SD-training   & 1.000 ± 0.000 & 1.000 ± 0.000 & 1.000 & 0.122 ± 0.173 & 0.250 ± 0.000 \\
& SD-sampling   & 0.896 ± 0.168 & 0.690 ± 0.456 & 0.735 & 0.100 ± 0.117 & 0.231 ± 0.063 \\
& AutoStop      & 0.88 5 ±  0.053 & 0.754 ± 0.174 & 0.912 & 0.115 ± 0.153 & 0.111 ± 0.056 \\
& IP-H       & 0.999 ± 0.002 & 0.035 ± 0.024 & 1.000 & 0.000 ± 0.000 & 0.249 ± 0.002 \\
			
\midrule
& OR            & 0.801 ± 0.000 & 0.026 ± 0.008 & 1.000 & 0.040 ± 0.000 & 0.001 ± 0.001 \\
& TM-adapted    & 0.862 ± 0.077 &  0.094 ± 0.049 &  1.000 &  0.022 ± 0.021 &  0.078 ± 0.097 \\
& QBCB          & 0.841 ± 0.006 &  0.070 ± 0.028 &  1.000 &  0.026 ± 0.002 &  0.051 ± 0.007 \\
& SCAL          & 0.039 ± 0.031 & 0.005 ± 0.001 & 0.000 & 0.924 ± 0.060 & 0.952 ± 0.039 \\
Legal & SD-training   & 1.000 ± 0.000 & 1.000 ± 0.000 & 1.000 & 0.007 ± 0.002 & 0.250 ± 0.000 \\
& SD-sampling   & 1.000 ± 0.000 & 1.000 ±  0.000 & 1.000 & 0.007 ± 0.002 & 0.250 ± 0.000 \\
& AutoStop      & 0.684 ± 0.022 & 0.794 ± 0.032 & 0.000 & 0.105 ± 0.015 & 0.145 ± 0.027 \\
& IP-H          & 0.793 ± 0.074 & 0.025 ± 0.000 & 0.500 & 0.045 ± 0.030 & 0.065 ± 0.012 \\
			
\bottomrule
\end{tabular}
\end{table}

\end{document}